%% file: main.tex
\documentclass{article}
\usepackage[margin=1in]{geometry} 
\usepackage{amsfonts,amssymb,amsmath,amsthm,mathtools}
\usepackage[ruled,linesnumbered,algoruled,boxed]{algorithm2e}
\usepackage{setspace}
\usepackage{enumitem}
\usepackage{caption}
\usepackage{subcaption}
\usepackage{pbox}

\usepackage[round]{natbib}
\usepackage{hyperref}
\hypersetup{colorlinks=true,linkcolor=red,citecolor=blue,urlcolor=black}
\usepackage{multirow,booktabs}
\usepackage{longtable}

\newtheorem{lemma}{Lemma}

\newtheorem{proposition}{Proposition}

\theoremstyle{definition}
\newtheorem{definition}{Definition}
\theoremstyle{remark}
\newtheorem{remark}{Remark}

\DeclareMathOperator*{\argmin}{arg~min~}

\newcommand{\F}{\text{F}} 
\newcommand{\I}{\mathrm{I}} 

\title{Structural Discovery with Partial Ordering Information \\for Time-Dependent Data with Convergence Guarantees}
\author{
Jiahe Lin\qquad 
Huitian Lei\qquad 
George Michailidis\footnote{Corresponding author. Department of Statistics, UCLA. \texttt{gmichail@ucla.edu}}
}

\date{}

\begin{document}

\maketitle
\input{00_abstract}
\input{01_Intro}

\input{02_Problem}
\input{03_Estimation}
\input{04_Simulation}
\input{05_Real}
\input{06_Conclusion}

\subsubsection*{Code Availability}
The code repository for this work is available at\\
\url{https://github.com/GeorgeMichailidis/high-dim-svar-partial-ordering}

\subsubsection*{Author Contribution}
JL proposed and implemented the method, conducted the experiments, compiled the results and drafted the manuscript. HL conducted the experiments partially and proofread the manuscript. GM conceived the project, contributed the proofs, drafted the manuscript and provided the computing resources.  

\vskip 0.5in
\setlength{\bibsep}{3pt}
\bibliographystyle{chicago}
\bibliography{main}

\clearpage
\appendix
\input{A1_admm_update}
\input{A2_algo_convergence}

\input{A3_background_id}
\input{A4_numerical}
\clearpage
\input{A5_listofVariables}

\end{document}

%% file: 00_abstract.tex
\begin{abstract}
Structural discovery amongst a set of variables is of interest in both static and dynamic settings. In the presence of lead-lag dependencies in the data, the dynamics of the system can be represented through a structural equation model (SEM) that simultaneously captures the contemporaneous and temporal relationships amongst the variables, with the former encoded through a directed acyclic graph (DAG) for model identification. 
In many real applications, a partial ordering amongst the nodes of the DAG is available, which makes it either beneficial or imperative to incorporate it as a constraint in the problem formulation. This paper develops an algorithm that can seamlessly incorporate a priori partial ordering information for solving a linear SEM (also known as Structural Vector Autoregression) under a high-dimensional setting. The proposed algorithm is provably convergent to a stationary point, and exhibits competitive performance on both synthetic and real data sets. 
\end{abstract}
{\it Keywords:} Structural Equation Models, Identifiability, Constrained Optimization, multi-block ADMM

%% file: 01_Intro.tex
\section{Introduction}\label{sec:intro}

Learning the interactions among a set of time series is a topic of interest and pertinent to applications in economics, functional genomics, neuroscience, environmental sciences and social media analysis (see, e.g., recent review papers \cite{runge2019inferring,vowels2022d} and references therein). Dynamic Bayesian Networks (DBN) capture conditional dependence relationships among a set of variables evolving over time and hence constitute a natural modeling framework for this task \citep{ghahramani1997learning}. They extend the notion of static graphical models over a set of variables as a function of time, wherein the structural relationships are encoded through a directed acyclic graph (DAG), and the temporal relationships are captured through lag dynamics. The problem of learning the parameters of DBNs from data has received significant attention in the literature; e.g., see \cite{scanagatta2019survey} and references therein. Further, when the relationships and dynamics are assumed to be linear, DBN can be expressed as a  {\em structural} Vector Autoregressive (SVAR) model (see also~\eqref{eqn:dgp}), that has been studied in the econometrics literature \citep{lutkepohl2005new,kilian2017structural}. However, the focus in the latter line of work has been on a small set of variables, whereas new application areas typically involve a large number of time series (i.e., of high dimension). Note that although the linearity assumption may occasionally be somewhat restrictive, linear models remain relevant and appealing in many real-world settings, due to its interpretability and parsimonious representation when used as a working model. 

In many applications, selected prior information is available for the structural relationships among variables. For example, in functional genomics there are known transcription factors that act only as regulators of other genes; analogous information is available for certain macroeconomic indicators, corresponding to the fact that as a group they cannot be descendent nodes to some other ones in a DAG. Therefore, it is important to incorporate such prior information in learning algorithms for the purpose of structural discovery. 

In this work, we develop an algorithm for estimating the parameters of large-scale SVAR models, which can incorporate the partial ordering information in a seamless way. 

\subsection{Related work}\label{sec:related-work}

We provide a brief review on existing approaches in the literature for learning the parameters of SVAR models. The key issue on the identifiability of its model parameters and the challenges it poses is presented in Section \ref{sec:formulation}.

\paragraph{Estimation of SVAR.} Classical work largely lies in the econometrics domain where methods have been developed primarily for fixed dimension SVAR models; e.g., \citet{stock2016dynamic,kilian2017structural,fry2011sign} and references therein. These methods ubiquitously start from the reduced VAR representation (see also equation~\eqref{eq:reduced-var}), then recover the structural parameter by imposing restrictions on the error covariance structure to achieve model identification. Recent developments on the topic amount to considering the structural component that captures variables' contemporaneous inter-dependencies as a DAG and perform ``causal" search or estimation. To that end, recent SVAR estimation methods are largely an extension of their respective DAG estimation counterparts, by considering a formulation that additionally incorporates the lag terms; estimation is done by either jointly considering the structural and the lag components or through a two-stage procedure that relies on residuals from the projection onto the lag space; e.g., \citet{hyvarinen2010estimation,moneta2011causal,malinsky2018causal,pamfil2020dynotears}. Some of these approaches can be extended to high dimensions.

\paragraph{Estimation of DAGs.} In light of the close connection between the SVAR and DAG problems, we briefly review approaches in estimating the latter next. There are three lines of work around this task: the first and most general one---in the absence of any additional prior information---include approaches that leverage greedy search algorithms over the space of the DAGs \citep{chickering2002learning,tsamardinos2006max}, those that rely on conditional independence tests \citep{spirtes2000causation,kalisch2007estimating} or likelihood-based ones~\citep{aragam2015concave,van2013ell1}. Note that the computational complexity of estimating a DAG from observational data is superexponential in the number of nodes/variables \citep{robinson1977counting} and thus many of them are greedy in nature and do not scale well even for moderate size problems involving 20-50 variables. More recently, optimization-based approaches \citep{zheng2018dags} and non-linear ones relying on neural networks have also been considered \citep{yu2019dag,lachapelle2019gradient}.
The second and rather restrictive approach, is that a {\em total topological ordering} for the nodes in the DAG $\mathcal{G}$ is known either from the literature or extensive experimental work on related settings (see discussion in \citet{markowetz2010understand} for functional genomics). The problem effectively boils down to estimating whether an edge is present as all potential parent nodes are known \citep{shojaie2010penalized}. The third and least explored category is to have limited information on the structural relationships among the variables in the form of a \textit{partial ordering} of the underlying nodes in $\mathcal{G}$; such information is available in a variety of applications and two examples are given in Section \ref{sec:realdata}. The notion of partial ordering will be formally defined in Section~\ref{sec:formulation}. 

In \citet{reisach2021beware}, the authors report that the performance of DAG estimation using continuous structural learning methods (or equivalently, optimization-based approaches; e.g., NOTEARS \citep{zheng2018dags}) can be sensitive to the data scale as measured by the concept of ``varsortability" introduced in that paper; specifically, these methods may face issues in the absence of high varsortability (i.e., when the marginal variance of the data is informative of the topological ordering). Given that selected recent methods for time-series data are built upon their DAG estimation counterparts (e.g., Dynotears \citep{pamfil2020dynotears} as an extension to NOTEARS), such susceptibility permeates. On the other hand, data in real-world applications may not possess strong varsortability, which may render estimates based on such methods unreliable. 

\paragraph{Contribution.} The main challenges in estimating high-dimensional SVAR models include identification of the model parameters and developing efficient algorithms for large-scale models. To this end, the key contribution of this paper is the development of a scalable and provably convergent algorithm to estimate the parameters of a SVAR model in a high dimensional regime. Additionally, the devised algorithm can seamlessly incorporate prior partial ordering information in the optimization problem formulation. Finally, note that despite being an optimization-based approach, the algorithm in this work does not face the same issue and is robust to data normalization. See in-depth discussion in Appendix~\ref{sec:varsortability}.

The remainder of the paper is organized as follows: Section~\ref{sec:formulation} gives the problem statement and discusses several key issues pertaining to the model in question, namely stability and model parameter identifiability. We present the proposed algorithm and briefly discuss its convergence property in Section~\ref{sec:method}, and assess its performance on synthetic and real datasets in Section~\ref{sec:simulation} and~\ref{sec:realdata}, respectively.

%% file: 02_Problem.tex
\section{Problem Statement}\label{sec:formulation}

Consider a system of $p$ variables $X_t :=(X_{t,1},\cdots,X_{t,p})^\top$ for which observations over time are collected. The dynamics of $X_t\in\mathbb{R}^p$ are assumed to be in accordance with the following SVAR with lag dynamics:
\begin{equation}\label{eqn:dgp}
    X_t = \mu + AX_{t} + B_1 X_{t-1} + \cdots + B_d X_{t-d} + \boldsymbol{\epsilon}_t, 
\end{equation}
wherein $A\in\mathbb{R}^{p\times p}$ captures the structural relationships among the $p$ variables, and $B_j\,(j=1,\cdots,d)$ the ``lead-lag" ones. It is further assumed that the error process $\boldsymbol{\epsilon}_t$ is independent and identically distributed across time points with mean zero and diagonal covariance matrix $\Sigma_{\boldsymbol{\epsilon}}$. In practical applications, $X_t$ usually has a non-zero mean and hence the SVAR model in~\eqref{eqn:dgp} would include an intercept term that can be estimated from the data (see also \citet{lutkepohl2005new}). Without loss of generality, we assume $X_t$ is mean-zero and omit the intercept term $\mu\in\mathbb{R}^p$ in~\eqref{eqn:dgp} in the remainder of this paper.

Next, we briefly elaborate on the issue of stability of the $X_t$ process and the identifiability of the model parameters, and also discuss how prior information on the structural relationships between the $X$ variables can be accommodated.

\paragraph{Stability of the process.} The SVAR model can be equivalently represented through a \textit{reduced} VAR(1) process as follows
\begin{equation}\label{eq:reduced-var}
\mathcal{X}_t = \Phi \mathcal{X}_{t-1} + v_t,  
\end{equation}
where 
$\mathcal{X}_t :=[X^\top_t, X^\top_{t-1}, \cdots, X^\top_{t-d+1}]^\top \in \mathbb{R}^{dp}$, $v_t := [u_t^\top, \mathbf{0}^\top,\cdots, \mathbf{0}^\top]^\top$ with $u_t:= (\mathrm{I}_p-A)^{-1}\boldsymbol{\epsilon}_t$; $\Phi$ is the transition matrix in the companion form, given by
{
\small
\begin{equation}\label{eq:Phi}
\Phi := \left[ \begin{smallmatrix}
(\mathrm{I}_p-A)^{-1} B_1 & \cdots & (\mathrm{I}_p-A)^{-1}  B_{d-1} & (\mathrm{I}_p-A)^{-1} B_d \\
\mathrm{I}_p & \cdots & O & O \\
\vdots & \ddots & \vdots & \vdots \\
O & \cdots & \mathrm{I}_p & O \end{smallmatrix}
\right] \in \mathbb{R}^{dp\times dp}.
\end{equation}
}%
A reduced VAR process is stable if $\text{det}(\mathrm{I}_{dp}-\Phi z)\neq 0$, for $z\leq 1$ \citep{lutkepohl2005new}. Based on standard results for reduced VAR processes \citep{basu2015regularized}, for $\{X_t\}$ to be stable (stationary), a sufficient condition is given by $\varrho(\Phi)<1$, with $\varrho(\cdot)$ denoting the spectral radius of a square matrix. Note that to obtain valid estimates of model parameters in~\eqref{eq:reduced-var} requires $X_t$ to be stable \citep{hamilton2020time}; hence the ensuing discussion on identifiability of model parameters is confined to such processes. 

\paragraph{Identifiability of model parameters.} The identification of model parameters $(A,B_1,\cdots,B_d)$ of the SVAR model is a key issue that has been extensively discussed in the literature. In particular, the difficulty stems from the contemporaneous dependency among the variables, as encoded by $A$, which requires at least $p(p+1)/2$ restrictions for it to be recovered if one starts from a reduced VAR representation. In this work, we assume that $A$ corresponds to the adjacency matrix of a directed acyclic graph (DAG), which is equivalent to the existence of some permutation(s) $\pi$ of the rows of $A$, such that $\pi(A)$ is a lower triangular matrix. Note that a restricted version of this assumption, namely imposing an a priori lower triangular structure to $A$ based on domain knowledge considerations, has been used in the econometrics literature for identification of fixed dimension SVAR models \citep{kilian2017structural,stock2016dynamic}.

On the other hand, in the absence of temporal dependency, a linear SEM of the form $X=AX+\boldsymbol{\epsilon}, X\in\mathbb{R}^p$ where $A$ encodes the underlying DAG $\mathcal{G}_A$, is not necessarily identifiable. The joint distribution $\mathbb{P}(X)$ of the observed variables is fully determined through the product distribution of the error variable $\mathbb{P}(\boldsymbol{\epsilon})$ and $\mathcal{G}_A$; conversely, however, the DAGs that give rise to the same $\mathbb{P}(X)$ are not unique \citep{spirtes2000causation}. In other words, multiple $\mathcal{G}_A$'s can be compatible with $\mathbb{P}(X)$ and thus the parameter $A$ is not uniquely identifiable from observational data without additional assumptions. For the purpose of identifiability, the following assumptions on the error distribution $\mathbb{P}(\epsilon)$ have been considered in the literature: (1) the distribution is non-Gaussian \citep{shimizu2006linear}; (2) the distribution is Gaussian with equal variance across its coordinates \citep{peters2014identifiability}; and (3) the distribution is Gaussian with unequal variances that are weakly monotonically increasing in the true ordering $\pi$ implied by the DAG $\mathcal{G}_A$ \citep{park2020identifiability}.

In summary, in this work, the identification scheme adopted for the parameters of the SVAR model in \eqref{eqn:dgp} encompasses the following assumptions: (a) $A$ is the adjacency matrix of a DAG, and (b) any of the above-mentioned three conditions on the distribution of the error $\mathbb{P}(\epsilon_t)$ hold.

\paragraph{Prior information and partial ordering.} As mentioned in Section~\ref{sec:intro}, in this work, the incorporation of prior information into the estimation procedure is enabled, with the former in the form of partial ordering. Formally, consider a (time-invariant) partition of the nodes $X_{t,1},\cdots,X_{t,p}$ into disjoint sets $\mathcal{V}_1,\cdots,\mathcal{V}_Q$, with $\mathcal{V}_1\prec \mathcal{V}_2\prec\cdots\prec\mathcal{V}_Q$, with $\prec$ denoting a precedence relationship, that is, there cannot be an edge $X_{t,i}\rightarrow X_{t,j}$ for $i\in \mathcal{V}_q,j\in \mathcal{V}_{q'},q'<q$. However, the intra-dependency or ordering of the variables within a set $\mathcal{V}_q, \forall \ q$ is not known and needs to be inferred from the data. In the extreme case where no prior information is available, the partition becomes trivial and all nodes effectively fall into one set.

%% file: 03_Estimation.tex
\section{A Provably Convergent Estimation Procedure}\label{sec:method}

For ease of exposition, in this section, we consider the special case where $d=1$ and let $B\equiv B_1$; the case where $d>1$ can be readily derived by stacking the lags and transition matrices which then gives the lag-1 representation (see representation in \eqref{eq:reduced-var} and \eqref{eq:Phi}).

To obtain estimates for model parameters $A$ and $B$, let $\mathbf{X}_n\in\mathbb{R}^{n\times p}$ denote the sample matrix with observations $\{x_1,\cdots,x_n\}$ stack in the rows of $\mathbf{X}_n$; $\mathbf{X}_{n-1}$ is analogously defined. The loss function is  
$\ell(A,B; \mathbf{X}_n,\mathbf{X}_{n-1}) := \frac{1}{2n}\|\mathbf{X}_n-\mathbf{X}_nA^\top - \mathbf{X}_{n-1}B^\top\|^2_{\F}$, and the optimization problem based on the $\ell_2$ loss is formulated as
\begin{equation}\label{eqn:solveA-B}
\small\setstretch{0.9}
\begin{split}
(\widehat{A},\widehat{B})~& :=\mathop{\text{argmin}}_{A,B} \Big\{ \ell(A,B; \mathbf{X}_n,\mathbf{X}_{n-1}) + \mu_A \|A\|_1 + \mu_B \|B\|_1 \Big\}, \\
& \text{subject to $A$ being acyclic},
\end{split}
\end{equation}
with the additional $\ell_1$-norm regularization terms inducing sparsity. In the presence of a {\em partial ordering} on $A$ (prior information), the search space of $A$ can be represented as
\begin{equation*}
\small
\begin{split}
&\mathcal{P}_A := \big\{ A\in\mathbb{R}^{p\times p}:~ A_{ij} = 0~\text{for}~(i,j)\in\mathcal{I}\times \mathcal{J}\big\},~~\text{where}~\mathcal{I}\times \mathcal{J}\subseteq \{1,\cdots,p\} \times \{1,\cdots,p\};
\end{split}
\end{equation*}
$A_{ij}=0\Leftrightarrow j\notin \text{pa}(i)$, that is, node $j$ cannot be a parent of node $i$ in the DAG representation. In the extreme case, $\mathcal{I}\times \mathcal{J}$ can be a null set, corresponding to the case where no prior information is available. As it can be seen later, our proposed algorithm can readily consume such partial ordering information and perform estimation in the restricted subspace $\mathcal{P}_A \subseteq \mathbb{R}^{p\times p}$.

\subsection{The proposed algorithm}

To solve ~\eqref{eqn:solveA-B}, we  leverage the results in~\citet{yuan2019constrained}, where acyclicity can be enforced through polyhedral constraints and the formulation can be solved via difference-convex (DC) programming and the augmented Lagrangian method of multipliers (ADMM). Concretely, Theorem~1 in \cite{yuan2019constrained} states that $A$ is acyclic, if and only if the following $p^3-p^2$ constraints are satisfied for some $\boldsymbol{\lambda}=[\lambda_{ij}] \in\mathbb{R}^{p\times p}$:
{\small
\begin{equation}\label{eqn:acyc}
    \lambda_{ik} + \mathbb{I}(j\neq k) - \lambda_{jk} \geq \mathbb{I}(A_{ij}\neq 0);~~i,j,k=1,\dots,p,i\neq j.
\end{equation}
}%
Together with the partial ordering information, by considering the truncated $\ell_1$-function $J_{\tau}(z):= \min(\frac{|z|}{\tau},1)$, $\tau\rightarrow 0$ as a surrogate for the indicator function, and introducing $\boldsymbol{\xi}=[\xi_{ijk}]\in\mathbb{R}^{p\times p\times p}, \xi_{ijk}\geq 0$ that translate inequality constraints to equality ones, the optimization problem can be written as 
\begin{equation}\label{eqn:solveA-B1}
\begin{split}
\mathop{\text{min}}_{A,B,\lambda}~~& \Big\{ \ell(A,B; \mathbf{X}_n,\mathbf{X}_{n-1}) + \mu_A \|A\|_1 + \mu_B \|B\|_1 \Big\},\\
\text{subject to}~~&\lambda_{ik} + \mathbb{I}(j\neq k) - \lambda_{jk} = J_{\tau}(A_{ij}) + \xi_{ijk},\\
& A\in\mathcal{P}_A, ~~~\xi_{ijk}\geq 0;~~~i,j,k=1,\dots,p,i\neq j.
\end{split}
\end{equation}
Note that in the case where $A_{ij}=0$ is a priori enforced, the corresponding constraint in~\eqref{eqn:acyc} can be simplified to $\lambda_{ik} + \mathbb{I}(j\neq k) - \lambda_{jk} = \xi_{ijk}$, which effectively is a relaxation to the right hand side for $(i,j)$. This is conceptually compatible with the nature of the acyclic constraint, that with $A_{ij}=0$, the graph could potentially accommodate other edges to be nonzero while still remains acyclic. Similar to~\citet{yuan2019constrained}, the formulation in~\eqref{eqn:solveA-B1} can be solved iteratively leveraging the decomposition $J_\tau(z)=|z|/\tau - \max\{|z|/\tau-1,0\}$ and the aid of an indicator matrix $\boldsymbol{w}\in\mathbb{R}^{p\times p}$, whose coordinates are given by $w_{ij} :=\mathbb{I}(|A_{ij}|<\tau)$, as outlined in Algorithm~\ref{algo:bigADMM}.
\begin{algorithm}[ht]
\small
\setstretch{0.8}
\caption{Solving for~\eqref{eqn:solveA-B1} via alternate update between $(A,B,\boldsymbol{\lambda})$ and $\boldsymbol{w}$.}
\KwIn{Data matrices $\mathbf{X}_n$, $\mathbf{X}_{n-1}$;  search space $\mathcal{P}_A$; and hyperparameters $\mu_A$, $\mu_B$ and $\tau$}\label{algo:bigADMM}
\While{not converging}{
$\bullet$ $\boldsymbol{w}$-update: update $\boldsymbol{w}$ according to~$w_{ij} \leftarrow \mathbb{I}(|A_{ij}| < \tau)$;\\
$\bullet$ $(A,B,\boldsymbol{\lambda})$-update: update $(A,B,\boldsymbol{\lambda})$ by solving for~\eqref{eqn:solveA-B2} with a fixed $\boldsymbol{w}$ via ADMM outlined in Algorithm~\ref{algo:ADMMupdate}.
\begin{equation}
\begin{split}
\mathop{\text{min}}\nolimits_{A,B,\lambda}~~& \{ \ell(A,B; \mathbf{X}_n,\mathbf{X}_{n-1}) + \mu_A \|A\|_1 + \mu_B \|B\|_1 \}, \label{eqn:solveA-B2}\\
\text{subject to}~~&\tau \lambda_{ik} + \tau\mathbb{I}(j\neq k) - \tau\lambda_{jk} = |A_{ij}|w_{ij} + \tau(1-w_{ij}) + \xi_{ijk},\\
&A\in\mathcal{P}_A, ~~~\xi_{ijk}\geq 0;~~~i,j,k=1,\dots,p,i\neq j.\\
\end{split}
\end{equation}
}
\KwOut{Estimated $\widehat{A}$ and $\widehat{B}$.}
\end{algorithm}

Next, we briefly outline the steps for the $(A,B,\boldsymbol{\lambda})$-update. To handle the non-differentiable parts of the objective function in~\eqref{eqn:solveA-B2}, we introduce $\widetilde{A}$ and $\widetilde{B}$, and the augmented Lagrangian function can be written as follows for some scaled variable $\rho>0$:
{\small
\begin{equation}\label{opt:L2}
\begin{split}
L_\rho(A,\widetilde{A},B,&\widetilde{B},\boldsymbol{\lambda},\boldsymbol{\xi};U_A, U_B, \boldsymbol{y})
:=~\ell(A,B; \mathbf{X}_n,\mathbf{X}_{n-1}) + \mu_A\|\widetilde{A}\|_1 + \mu_B \|\widetilde{B}\|_1 \\
& + \frac{\rho}{2}\|A - \widetilde{A}\|_{\F}^2 + \rho\langle A-\widetilde{A}, U_A\rangle + \frac{\rho}{2}\|B - \widetilde{B}\|_{\F}^2  + \rho\langle B-\widetilde{B}, U_B\rangle + \mathcal{T}_1 + \mathcal{T}_2,
\end{split}
\end{equation}
}%
with
{\small
\begin{align*}
\mathcal{T}_1 &:=~\frac{\rho}{2}\sum\nolimits_k\sum\nolimits_{i \neq j}\Big( |\widetilde{A}_{ij}|w_{ij} + \tau(1-w_{ij}) + \xi_{ijk} - \tau \lambda_{ik} - \tau \mathbb{I}(j\neq k) + \tau\lambda_{jk}\Big)^2, \\
\mathcal{T}_2& :=~\rho\sum\nolimits_k\sum\nolimits_{i \neq j}y_{ijk}\Big( |\widetilde{A}_{ij}|w_{ij} + \tau(1-w_{ij}) + \xi_{ijk} - \tau \lambda_{ik} - \tau \mathbb{I}(j\neq k) + \tau\lambda_{jk}\Big);
\end{align*}
}%
$U_A, U_B, \boldsymbol{y}$ are dual variable matrices/tensors. One proceeds with primal descent on $(A,\widetilde{A},B,\widetilde{B},\boldsymbol{\lambda},\boldsymbol{\xi})$ and dual ascent on $(U_A,U_B,\boldsymbol{y})$ as outlined in Algorithm~\ref{algo:ADMMupdate}; see Appendix~\ref{appendix:solveADMM} for the exact update of each step. 
\begin{algorithm}[htb]
\small
\setstretch{0.7}
\SetAlgoLined
\caption{Update $(A,B,\boldsymbol{\lambda})$ via multi-block ADMM: a schematic outline}\label{algo:ADMMupdate}
\KwIn{Data matrices $\mathbf{X}_n$, $\mathbf{X}_{n-1}$; search space $\mathcal{P}_A$; fixed $\boldsymbol{w}$; hyperparameters $\mu_A$, $\mu_B$, $\tau$ and $\rho$}
\BlankLine
\While{not converging}{
\BlankLine
{\bf(Primal descent)}\\
$\bullet$ cyclic update on blocks $A$, $\widetilde{A}$, $B$, $\widetilde{B}$, $\boldsymbol{\lambda}$, $\boldsymbol{\xi}$ by minimizing~\eqref{opt:L2} w.r.t. the block of interest;
\BlankLine
{\bf(Dual ascent)}\\
$\bullet$ $U_A$-update: $U_A^{(s)} \leftarrow U_A^{(s-1)} + (A^{(s)}-\widetilde{A}^{(s)})$\\
$\bullet$ $U_B$-update: $U_B^{(s)} \leftarrow U_B^{(s-1)} + (B^{(s)}-\widetilde{B}^{(s)})$\\
$\bullet$ $\boldsymbol{y}$-update: $y^{(s)}_{ijk}  \leftarrow  y^{(s-1)}_{ijk} + (|\widetilde{A}_{ij}^{(s)}|w_{ij} + \tau(1-w_{ij}) + \xi^{(s)}_{ijk} - \tau \lambda_{ik}^{(s)}-\tau \mathbb{I}(j\neq k) +\tau \lambda_{jk}^{(s)})$
}
\KwOut{Solution $\widehat{A}$ and $\widehat{B}$ to~\eqref{opt:L2}}
\end{algorithm}
It is worth noting that given the specific form of the augmented Lagrangian, all primal updates possess closed-form minimizers, which empirically aids in fast and stable convergence of the cyclic block-updates. 

To conclude this subsection, we briefly comment on how the partial ordering information is incorporated through block updates. First, note that by introducing $\widetilde{A}$ that separates the non-differentiable part of $A$, after some algebra, the update of $A$ (while holding $\widetilde{A},U_A,B$ fixed) can be written as
{\small
\begin{align*}
A \leftarrow \argmin\limits_{A\in\mathcal{P}_A} \text{trace}\Big\{ \frac{1}{2} A\big[(\frac{1}{n}\mathbf{X}_n^\top\mathbf{X}_n) + \rho \mathrm{I}_p\big]A^\top 
- \big[ (\frac{1}{n}\mathbf{V}^\top_n\mathbf{X}_n) + \rho(\widetilde{A} - U_A) \big] A^\top \Big\},
\end{align*}
}%
where $\mathbf{V}_n := \mathbf{X}_n-\mathbf{X}_{n-1}B^\top$.
The update is separable for each row of $A$; additionally, the prior partial ordering information---in the form of restricting the skeleton indices of each row of $A$ to a subset of $\{1,\cdots,p\}$---becomes equivalent to considering only the corresponding column sub-space of the design matrix.

\subsection{Convergence analysis}

We provide a brief discussion on the convergence property of the proposed algorithm, while deferring all lemmas and their proofs to Appendix~\ref{appendix:algo_conv}. 

Note that the $\boldsymbol{w}$-update step is straightforward, which effectively boils down to obtaining the complement of the support of the estimated $A$ at each iteration. The ensuing analysis establishes convergence properties of Algorithm \ref{algo:ADMMupdate}.

Denote by $\Theta :=(A,\widetilde{A},B,\widetilde{B},\boldsymbol{\lambda},\boldsymbol{\xi})$ and $\Psi :=(U_A,U_B,\boldsymbol{y})$ the collection of primal and dual variables, respectively. 

\begin{proposition}\label{prop:convergence} Consider a sequence of iterates $(\Theta^{(s)},\Psi^{(s)})$
generated by Algorithm ~\ref{algo:ADMMupdate}, indexed by $s$. Then, the sequence converges to a stationary point of the augmented Lagrangian function~\eqref{opt:L2} for any initial point $(\Theta^{(0)},\Psi^{(0)})$.
\end{proposition}

We provide some insights on the critical steps required to achieve convergence. The first is a ``sufficient descent property"; namely, one needs to find a positive constant $\eta$ so that two successive iterates of the primal and dual variables satisfy $\eta\|(\Theta^{(s+1)},\Psi^{(s+1)})-(\Theta^{(s)},\Psi^{(s)})\|_{\F}^2\leq L_\rho(\Theta^{(s)},\Psi^{(s)})-L_\rho(\Theta^{(s+1)},\Psi^{(s+1)}),~s=0,1,...$. This is established in Lemma~\ref{lemma-suff-dec}. The second is a subgradient lower bound for the gap between successive iterates; namely,
there exists another positive constant $\gamma$ such that
any element $C^{(s)}$ in the subdifferential of $L_\rho(\Theta^{(s)},\Psi^{(s)})$ satisfies
$\|C^{(s+1)}\|_{\F}^2\leq \gamma \|(\Theta^{s+1)},\Psi^{(s+1)})-(\Theta^{(s)},\Psi^{(s)})\|_{\F}^2$. This is established in Lemma~\ref{lemma-limit-point} with the aid of Lemma~\ref{lemma-subsequence}. Note that these two requirements are satisfied by most ``good" descent algorithms. Further, when the above two properties hold, then the accumulation points of \textit{any} algorithm is a non-empty, compact and connected set (see Remark 5 in \citet{bolte2014proximal}).
However, the existence of $\eta,\gamma$ \textit{depend on the structure of the specific algorithm used}; the results in Lemmas~\ref{lemma-sandwich} and \ref{lemma-limit-point} show how to obtain them given the structure and updates of the developed multi-block ADMM in Algorithm~\ref{algo:ADMMupdate}. The last requirement to establish global-convergence-to-a-critical-point of $L_\rho(\cdot)$ does not depend on the structure of the algorithm used, but on the type of function $L_\rho(\cdot)$. To that end, Lemma~\ref{lemma-KL} shows that the augmented Lagrangian satisfies the Kurdyka-\L{}ojasiewicz property \citep{kurdyka1998gradients}, which aids in establishing that the sequence of iterates $(\Theta^{(s)},\Psi^{(s)})$ generated by Algorithm~\ref{algo:ADMMupdate} is a Cauchy sequence.

\begin{proof}[Proof of Proposition~\ref{prop:convergence}]
Based on the results of Lemmas~\ref{lemma-subsequence} and~\ref{lemma-limit-point}, we have that $(\Theta^{(s)},\Psi^{(s)})$ is a bounded sequence and the set of limit points of $(\Theta^{(s)},\Psi^{(s)})$ when initialized at $(\Theta^{(0)},\Psi^{(0)})$ is non-empty, respectively. Further, existing results in the literature---in particular, Lemma~\ref{lemma-limit-point} and Remark~5 in \cite{bolte2014proximal}---ensure the compactness of the set of limit points of the sequence $(\Theta^{(s)},\Psi^{(s)})$, when the latter is initialized at $(\Theta^{(0)},\Psi^{(0)})$. The remainder of the proof follows along the lines of Theorem~1 in \cite{bolte2014proximal} by utilizing the Kurdyka-\L{}ojasiewicz property of the augmented Lagrangian function, as established in Lemma~\ref{lemma-KL}. 
\end{proof}

\begin{remark} The class of functions that satisfy the Kurduka-\L{}ojasiewicz property is remarkably large and includes many loss functions, regularization terms, as well as polyhedral constraints used in machine learning tasks. Further, the proof strategy is applicable to many algorithms. In this paper, we provide the details for a multi-block ADMM algorithm for the non-convex, non-smooth problem arising from the SVAR problem formulation under consideration. Note that such algorithms are used in many other machine learning problems sharing similar features and hence the proof is of general interest. Finally, note that the proposed algorithm exhibits global convergence to a critical point, i.e., such convergence is independent of the algorithm's initialization, which is an attractive feature in practical applications.
\end{remark}
Empirically, the proposed algorithm exhibits stable convergence; the alternating update between $\boldsymbol{w}$ and $(A,B,\boldsymbol{\lambda})$ usually converges within 10 iterations; the $(A,B,\boldsymbol{\lambda})$-update step that relies on ADMM typically converges within 100 iterations, although during the very first round of the outer update, it often requires more. 

\begin{remark}
\cite{yuan2019constrained} provide a brief proof for the ADMM-based algorithm developed for reconstructing DAG from iid data. Specifically, the proof assumes that the augmented Lagrangian function is \textit{strongly convex} and appeals to a result in \cite{boyd2011distributed} to establish convergence of the algorithm. However, note that the augmented Lagrangian function is not strongly convex; additionally, the result in \cite{boyd2011distributed} only holds for a two-block ADMM algorithm, rather than the multi-block updates used in \cite{yuan2019constrained} and the current work. Indeed, establishing convergence for multi-block ADMM even for convex functions was challenging and remained open for awhile, as attested in \cite{chen2016direct}. In this work, the proof of Proposition~\ref{prop:convergence} takes a different route and leverages a road map outlined in \cite{bolte2014proximal} that establishes the convergence of ``descent-type algorithms" and only requires the Kurdyka-\L{}ojasiewicz property of the augmented Lagrangian function, which is significantly weaker.
\end{remark}

%% file: 04_Simulation.tex
\section{Synthetic Data Experiments}\label{sec:simulation}

We evaluate the performance of the proposed algorithm and the effectiveness of incorporating the partial ordering information as priors in the estimation through a series of synthetic data experiments. 

\paragraph{Settings.} The data are generated according to an SVAR model with $d=2$ lags, that is,
\begin{align} 
    &\quad X_t = A X_{t} + B_1 X_{t-1} + B_2 X_{t-2} + \boldsymbol{\epsilon}_t,\label{eqn:simdgp}\\
    &\text{where}~~\mathbb{E}(\boldsymbol{\epsilon}_t) = 0,~\Sigma_{\boldsymbol{\epsilon}} := \text{Cov}(\boldsymbol{\epsilon}_t) = \text{diag} \big(\sigma_1^1,\cdots, \sigma_p^2\big); \nonumber
\end{align}
coordinates of the noise component are independent and potentially heteroscedastic, depending on the distribution from which it is drawn. We consider cases where the system consists of 100 variables, with the structural parameter $A$ exhibiting varying degree of sparsity and the noise component $\boldsymbol{\epsilon}_t$ drawn from different distributions; see Table~\ref{tab:simsetting}. \footnote{Recall that as discussed in Section~\ref{sec:formulation}, different sets of assumptions have been provided in the literature to guarantee the identifiability of the underlying DAG in the SVAR model. In our experiment setup, we consider settings where the error distribution is either Gaussian with unequal variances that are weakly monotonically increasing \citep{park2020identifiability}, or non-Gaussian \citet{shimizu2006linear}; as such, they respectively satisfy assumptions (3) and (1) in the aforementioned discussion.}

\begin{table*}[ht]
\scriptsize
\setstretch{0.9}
\centering
\caption{Parameter setup for synthetic data experiments. $s_{\cdot}$ denotes the sparsity level of the corresponding parameter; $(l_{\cdot},u_{\cdot})$ corresponds to the lower and upper bounds of the (initial) draws of the signals; $\sigma_i$ corresponds to the standard deviation of the coordinates of the noise component.}
\label{tab:simsetting}
\begin{tabular}{r|cccccccl}
\hline
setting id  & $p$ & $s_A$ & $(l_A,u_A)$ & $s_{B_1}$ & $s_{B_2}$ & $(l_B,u_B)$ & $\sigma_i$ & noise dist \\  \hline
S1 &  100 & 0.05 & $(0.25,0.9)$ & 0.05 & 0.02 & $(1,3)$ & $\mathsf{Unif}[0.8,2]$ & Gaussian \\ 
S2 &  100 & 0.10 & $(0.25,0.7)$ & 0.05 & 0.02 & $(1,3)$ & $\mathsf{Unif}[0.8,2]$ & Gaussian  \\
S3 &  100 & 0.05 & $(0.25,0.9)$ & 0.05 & 0.02 & $(1,3)$ & $1$ & Laplace \\
S4 &  100 & 0.10 &  $(0.25,0.7)$ & 0.05 & 0.02 & $(1,3)$ & $1$ & Laplace \\
S5 &  100 & 0.05 & $(0.25,0.9)$ & 0.05 & 0.02 & $(1,3)$ & $1$ & Student's-t (df=4) \\
S6 &  100 & 0.10 & $(0.25,0.7)$ & 0.05 & 0.02 & $(1,3)$ & $1$ & Student's-t (df=4) \\
\hline
\end{tabular}
\end{table*}
Note that to ensure the stability of the process, the spectral radius $\varrho$ of the companion matrix for the corresponding reduced VAR, that is,
{\small
\begin{equation*}
    \Phi(A,B_1,B_2) := \begin{bmatrix}
(\mathrm{I}_p-A)^{-1} B_1 & (\mathrm{I}_p-A)^{-1}  B_2 \\
\mathrm{I}_p & O \end{bmatrix}
\end{equation*}
}%
needs to be strictly less than 1 (see also Section~\ref{sec:formulation}); to achieve this, we proceed as in the following steps: 
\begin{enumerate}[itemsep=0pt,topsep=1pt]
\item For transition matrices $B_1$ and $B_2$, their skeletons are determined by independent draws from $\mathsf{Bernoulli}(s_{B_1})$ and $\mathsf{Bernoulli}(s_{B_2})$, respectively; nonzero entries are first drawn from $\pm\mathsf{Unif}(l_B,u_B)$, then scaled such that $\varrho\big(\Phi(O,B_1,B_2)\big) = 0.5$\footnote{Here we set 0.5 as the target spectral radius; however it typically cannot be attained exactly except for $\text{VAR}(1)$.}, where $\Phi(O,B_1,B_2)$ corresponds to the companion matrix of the reduced VAR if one ignores the structural component.
\item For the structural parameter $A$, to obtain its skeleton subject to the acyclic constraint, each entry in the lower diagonal is drawn independently from $\mathsf{Bernoulli}(s_A)$; nonzero entries are then drawn from $\pm\mathsf{Unif}(l_A,u_A)$.
\item Repeat steps 1 and 2 if $\varrho\big(\Phi(A,B_1,B_2)\big)<1$ is not satisfied.
\end{enumerate}
In practice, the above procedure gives a set of parameters that yield $\varrho(\Phi(A,B_1,B_2))\approx 0.95$ within a few trials. A smaller spectral radius can be attained, if one further reduces the signal strength. Once model parameters are generated, we generate $\{X_t\}$ according to~\eqref{eqn:simdgp}; the $\boldsymbol{\epsilon}_t$'s are either Gaussian (S1, S2) or Laplace distributed (S3, S4): in the former case, the $\sigma_i$'s for each coordinate $i=1,\cdots,p$ are drawn from $\mathsf{Unif}(0.8,2)$ then sorted according to the topological ordering of the nodes as dictated by $A$; in the latter case, $\sigma_i\equiv 1$ . In settings S5 and S6, we additionally consider the case where the noise are generated from $t$-distribution, to test the robustness of the proposed method in the presence of heavy tails.

For all settings, we run the proposed algorithm on data with sample sizes $n=50, 100, 200$ and a varying level of available prior information provided through a partial ordering constraint, that is, 10\%, 20\%, 50\% of the complement of the support set, i.e., $\{(i,j): A_{ij}=0; (i,j)\in\{1,\cdots,p\}\times\{1,\cdots,p\}\}$. Note that the case with $n=50$ is a rather challenging setting: considering the number of parameters to be estimated, the estimation is ``under-powered".  

\begin{remark}\label{rmk:density} We briefly comment on the sparsity level adopted in the experiment settings. Consider a limiting case where a total topological ordering of the nodes is known a priori; the DAG learning problem reduces to selecting the parent node set by using sparse regression techniques \citep{reisach2021beware}. For i.i.d data, under high dimensional scaling, the sparsity level allowed for consistent estimation of the skeleton is $s\sim o\big(\tfrac{n}{\log(p^2)}\big)$. Further, in the SVAR setting where the data exhibit temporal dependence, the sparsity level is impacted by an additional $\kappa$ factor that quantifies the temporal dependence, that is: $s\sim o\big(\tfrac{n}{\kappa^2 \log(p^2)}\big)$, where  $\kappa=\tfrac{M}{m}>1$ with $M$ and $m$ denoting the maximum and minimum eigenvalues of the spectral density of the time series data under consideration, respectively \citep[see, e.g.,][for sub-Gaussian errors]{basu2015regularized}. Based on the above, the sparsity considered in the above settings is fairly high, even if a total topological ordering were given. In our experiments, at most some partial topological ordering information is available; consequently, the permissible level of sparsity  further reduces when compared to the limiting case.
\end{remark}

\paragraph{Performance evaluation.} We focus assessment on the structural component $A$, and specifically on skeleton recovery across different model settings, sample sizes and percentage of prior information provided, as shown in Table~\ref{tab:simres}. Results on the lag-components $B$ and the overall goodness-of-fit of the algorithm are provided in Appendix~\ref{appendix:lag}. In particular, to understand the impact of incorporating prior information, we report the True Positive Rate (TP, or recall, equivalently) and True Negative Rate (TN, or $1-\text{false positive rate}$, equivalently) for support recovery over different sample sizes and prior setups. $\{00,10,20,50\}$ correspond to varying level of partial ordering information---from no prior (00) to $50\%$ (50) of the non-support---given as a prior constraint. 
\begin{table*}[h]
\scriptsize
\setstretch{0.9}
\centering
\caption{Evaluation for $\widehat{A}$ obtained using \textbf{our proposed method}: Results are based on the median of 10 replicates, with the standard deviation of the corresponding metric reported in parentheses.}\label{tab:simres}
\begin{tabular}{rr|cc|cc|cc|cc}
\toprule
&  & \multicolumn{2}{c|}{00} &  \multicolumn{2}{c|}{10} & \multicolumn{2}{c|}{20} & \multicolumn{2}{c}{50} \\ \cmidrule(lr){3-4} \cmidrule(lr){5-6} \cmidrule(lr){7-8} \cmidrule(lr){9-10}
& $n$ & TP & TN & TP & TN & TP & TN & TP & TN\\ \hline

S1 & 50  & 0.69(0.040) & 0.79(0.006) & 0.69(0.051) & 0.81(0.006) & 0.65(0.042) & 0.82(0.005) & 0.80(0.030) & 0.85(0.006) \\
   & 100 & 0.78(0.020) & 0.85(0.004) & 0.78(0.020) & 0.87(0.004) & 0.77(0.030) & 0.88(0.005) & 0.86(0.025) & 0.91(0.006) \\
   & 200 & 0.88(0.018) & 0.88(0.005) & 0.87(0.017) & 0.89(0.006) & 0.87(0.017) & 0.90(0.006) & 0.95(0.014) & 0.93(0.006) \\\midrule

S2 & 50  & 0.56(0.056) & 0.72(0.004) & 0.53(0.059) & 0.74(0.003) & 0.56(0.035) & 0.76(0.003) & 0.73(0.031) & 0.77(0.009) \\
   & 100 & 0.73(0.046) & 0.74(0.011) & 0.71(0.031) & 0.76(0.007) & 0.73(0.030) & 0.78(0.013) & 0.84(0.026) & 0.81(0.009) \\
   & 200 & 0.84(0.014) & 0.83(0.004) & 0.83(0.015) & 0.84(0.005) & 0.84(0.015) & 0.86(0.003) & 0.93(0.011) & 0.89(0.006) \\\midrule

S3 & 50  & 0.69(0.035) & 0.84(0.007) & 0.69(0.041) & 0.86(0.006) & 0.70(0.026) & 0.87(0.005) & 0.83(0.031) & 0.89(0.005) \\
   & 100 & 0.84(0.015) & 0.86(0.005) & 0.84(0.017) & 0.88(0.005) & 0.83(0.020) & 0.89(0.004) & 0.93(0.020) & 0.91(0.005) \\
   & 200 & 0.91(0.011) & 0.91(0.002) & 0.91(0.012) & 0.92(0.002) & 0.92(0.012) & 0.92(0.003) & 0.98(0.008) & 0.95(0.004) \\\midrule

S4 & 50  & 0.61(0.067) & 0.79(0.012) & 0.53(0.043) & 0.80(0.008) & 0.52(0.031) & 0.81(0.005) & 0.73(0.041) & 0.84(0.005) \\
   & 100 & 0.77(0.044) & 0.78(0.014) & 0.77(0.036) & 0.79(0.009) & 0.77(0.031) & 0.81(0.007) & 0.88(0.032) & 0.83(0.009) \\
   & 200 & 0.85(0.014) & 0.84(0.006) & 0.86(0.009) & 0.85(0.004) & 0.87(0.010) & 0.86(0.005) & 0.95(0.010) & 0.90(0.007) \\\midrule

S5 & 50  & 0.68(0.034) & 0.86(0.004) & 0.69(0.029) & 0.87(0.005) & 0.69(0.033) & 0.88(0.005) & 0.76(0.033) & 0.90(0.006) \\
   & 100 & 0.85(0.044) & 0.84(0.008) & 0.86(0.044) & 0.85(0.006) & 0.86(0.035) & 0.86(0.004) & 0.91(0.021) & 0.89(0.006) \\
   & 200 & 0.89(0.012) & 0.86(0.007) & 0.89(0.012) & 0.87(0.005) & 0.89(0.017) & 0.88(0.005) & 0.97(0.010) & 0.91(0.003) \\\midrule

S6 & 50  & 0.61(0.042) & 0.76(0.005) & 0.60(0.045) & 0.78(0.004) & 0.63(0.033) & 0.80(0.005) & 0.74(0.055) & 0.81(0.005) \\
   & 100 & 0.79(0.067) & 0.77(0.010) & 0.79(0.037) & 0.78(0.005) & 0.80(0.058) & 0.80(0.009) & 0.81(0.042) & 0.81(0.008) \\
   & 200 & 0.85(0.013) & 0.84(0.005) & 0.85(0.015) & 0.85(0.004) & 0.86(0.014) & 0.86(0.004) & 0.95(0.006) & 0.90(0.004) \\
\bottomrule
    \end{tabular}
\end{table*}
Based on the results in Table~\ref{tab:simres}, the main findings are threefold: (1) despite similar setups for all other model parameters, model performance is superior in the case where the noise distribution is Laplace/Student's-t compared to Gaussian with monotonically increasing variances, provided that the estimation is moderately powered (e.g., $n=100,200$), as manifested by a higher detection of the true skeleton (i.e., true positive rate) without compromising the true negatives; (2) Although the prior partial ordering information is in the form of zero-constraints, it imposes restrictions on the search space of the skeleton and is integrated throughout the estimation process; therefore, the benefit of incorporating prior information is not limited to ruling out false positives, but it can also promote discovery. (3) As one would expect, the estimation becomes more challenging as the graph becomes more dense, as manifested by significantly lower true positive rate, especially for low sample size settings ($n=50,100$). Finally, note that the methodology is robust to the presence of heavy tails, as manifested by the overall comparable performance across settings with different noise distributions, provided all else held identical.

The performance of our proposed algorithm is also benchmarked against SVAR-GFCI \citep{malinsky2018causal} for settings S1-S4, with the latter being a score-based method that uses greedy optimization on the model score to learn the graph, followed by carrying out statistical tests for conditional independence to orient the edges; see Table~\ref{tab:simres-tetrad}. In particular, we leverage the python implementation of TETRAD\footnote{\url{https://github.com/cmu-phil/py-tetrad}}, wherein the available prior information can be passed in as an argument. 

\begin{table*}[h]
\scriptsize
\setstretch{0.9}
\centering
\caption{Evaluation for $\widehat{A}$ obtained using \textbf{SVAR-GFCI}. Results are based on the median of 10 replicates, with the standard deviation of the corresponding metric reported in parentheses.}\label{tab:simres-tetrad}
\begin{tabular}{rr|cc|cc|cc|cc}
\toprule
&  & \multicolumn{2}{c|}{00} &  \multicolumn{2}{c|}{10} & \multicolumn{2}{c|}{20} & \multicolumn{2}{c}{50} \\ \cmidrule(lr){3-4} \cmidrule(lr){5-6} \cmidrule(lr){7-8} \cmidrule(lr){9-10}
& $n$ & TP & TN & TP & TN & TP & TN & TP & TN\\ \hline
S1 & 50 & 0.21(.03) & 0.99(.001) & 0.22(.03) & 0.99(.001) & 0.22(.03) & 1.00(.001) & 0.30(.03) & 1.00(.001) \\
& 100 & 0.28(.03) & 0.99(.001) & 0.31(.03) & 0.99(.001) & 0.33(.03) & 1.00(.001) & 0.45(.03) & 1.00(.001) \\
& 200 & 0.32(.05) & 0.99(.001) & 0.35(.04) & 0.99(.001) & 0.37(.04) & 1.00(.001) & 0.57(.03) & 1.00(.001) \\ \midrule
S2 & 50 & 0.06(.01) & 0.99(.001) & 0.07(.01) & 0.99(.001) & 0.08(.01) & 0.99(.001) & 0.11(.03) & 0.99(.001) \\
& 100 & 0.09(.01) & 0.99(.001) & 0.10(.01) & 0.99(.001) & 0.11(.01) & 0.99(.001) & 0.16(.03) & 0.99(.001) \\
& 200 & 0.09(.01) & 0.99(.001) & 0.10(.01) & 0.99(.001) & 0.11(.01) & 0.99(.001) & 0.16(.02) & 0.99(.001) \\ \midrule
S3 & 50 & 0.20(.03) & 0.99(.001) & 0.21(.03) & 0.99(.001) & 0.22(.03) & 1.00(.001) & 0.30(.03) & 1.00(.001) \\
& 100 & 0.28(.03) & 0.99(.001) & 0.30(.03) & 0.99(.001) & 0.31(.03) & 0.99(.001) & 0.47(.03) & 1.00(.001) \\
& 200 & 0.35(.05) & 0.99(.001) & 0.37(.04) & 0.99(.001) & 0.41(.04) & 1.00(.001) & 0.58(.03) & 1.00(.001) \\ \midrule
S4 & 50 & 0.06(.01) & 0.99(.001) & 0.07(.01) & 0.99(.001) & 0.08(.03) & 0.99(.001) & 0.12(.03) & 0.99(.001) \\
& 100 & 0.08(.01) & 0.99(.001) & 0.10(.01) & 0.99(.001) & 0.12(.01) & 0.99(.001) & 0.15(.03) & 0.99(.001) \\
& 200 & 0.07(.01) & 0.99(.001) & 0.08(.01) & 0.99(.001) & 0.10(.01) & 0.99(.001) & 0.16(.02) & 0.99(.001) \\
\bottomrule
\end{tabular}
\end{table*}
A major issue with SVAR-GFCI observed under the settings in consideration is its low discovery rate; this may be due to high dimensionality and low sample size, and it is more pronounced for denser graphs. The partial ordering information aids in improved discovery of the graph skeletons, as what one would expect. Further, contrary to the estimates obtained using our proposed method, here we do not observe discrepancy in terms of recovery performance between the two cases, where the noise distribution is Gaussian versus being Laplace. 

Comparison with several other methods \citep[e.g.,][]{pamfil2020dynotears,hyvarinen2010estimation} whose existing implementation does not readily consume prior information is deferred to Appendix~\ref{appendix:sim}, where the comparison is only conducted for the case without partial ordering. Additionally, we also include a discussion related to varsortability \citep{reisach2021beware} and additional results to display the impact from data normalization in Appendix~\ref{sec:varsortability}.

%% file: 05_Real.tex
\section{Real Data Analysis}\label{sec:realdata}

To evaluate how our proposed algorithm would perform in real world settings, we consider two applications and examine the structural and temporal components estimated from the proposed method. 

\subsection{US macroeconomic data}

SVAR models are widely used to address various problems in macroeconomic analysis, including the effect of monetary interventions by central banks to the economy \citep{christiano2005nominal}. However, small VAR models regularly used in such analyses lead to empirical results that may be contradictory to economic theory tenets \citep{sims1980macroeconomics}. It has been suggested that large scale SVAR models could overcome such difficulties, however, their identification is typically challenging. The proposed approach offers a principled strategy to use large SVARs. 

The dataset comprises of a number of US macroeconomic indicators measured at quarterly frequency, sourced from the FRED-QD database \citep{mccracken2020fred}. We consider 78 variables spanning the period from 1Q1973 to 2Q2022, totaling 200 observations. These variables encompass several different categories and capture different facets of the economy, with the major ones being industrial production, producer and consumer price indices and their components (tier 1, low tier), Federal Funds Rate (FFR) as an approximation of monetary policy (tier 2, intermediate tier), and several market variables including treasury yields, S\&P500 and NASDAQ composite indices (tier 3, high tier); see \citet{bernanke2005measuring} and discussion therein for inclusion of variables in the model. A prior partial ordering is constructed based on variables' tiers; in particular, we do not allow variables from a higher tier to be the parent nodes of those in a lower tier, that is, the following directional relationship is {\em prohibited}: \{(tier 2, tier 3)~$\rightarrow$ tier 1; tier 3~$\rightarrow$ tier 1\}. This is predicated on the premise that tier 1 variables are ``slow moving" and hence not impacted within the same time period by the FFR (tier 2) or ``fast moving" variables in tier 3, the latter being sensitive to contemporaneous economic information and shocks \citep{bernanke2005measuring}. Finally, to ensure stationarity of the time series, we apply the benchmark transformation suggested in \citet{mccracken2020fred}, which follows from \citet{stock2012disentangling,stock2012generalized}; further, these time series are de-meaned before they are fed into the model.

We run the proposed algorithm on this dataset, with the hyperparameters $\mu_A,\mu_B$ selected so that the one-step-ahead predictive RMSE is minimized; recall, that the SVAR model can be expressed in a reduced form as in~\eqref{eq:reduced-var} which gives the recursive relationship for prediction. We set the number of lags $d=2$, trying to strike a balance between parsimony by not over-expanding the model parameter space and capturing delayed effects through adequate inclusion of lags. The obtained results show that with $d=2$, the magnitude of the estimated parameters in $B_2$ is getting significantly smaller than those in $B_1$. 

\begin{figure*}
    \centering
    \begin{subfigure}[t]{0.45\textwidth}
         \centering
    \includegraphics[scale=0.38]{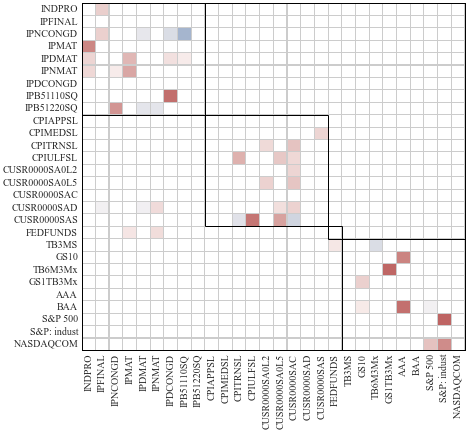}
    \caption{A partial view of the estimated $\widehat{A}$ (structural parameter) encompassing industrial production and consumer price indices (tier 1), FFR (tier 2) and market variables (tier 3). }\label{fig:heatA}
    \end{subfigure}
     \qquad
     \begin{subfigure}[t]{0.45\textwidth}
     \centering
    \includegraphics[scale=0.38]{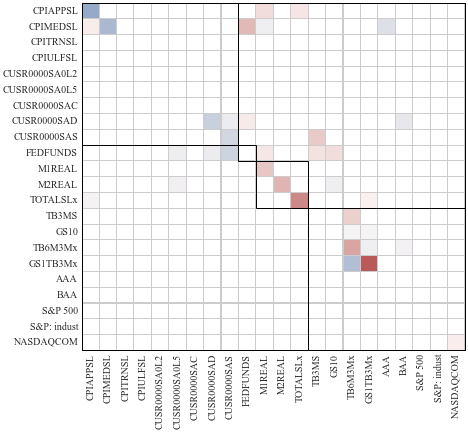}
    \caption{A partial view of the estimated $\widehat{B}_1$ (lag parameter) encompassing consumer price indices, FFR, money stock and market variables.}\label{fig:heatB1}
     \end{subfigure}
\caption{Partial views of the estimated structural parameter $A$ and lag parameter $B_1$. Parent variables are depicted in the columns and their descendants are in the rows.}
\end{figure*}

The heatmap in Figure~\ref{fig:heatA} shows the impact of industrial production and consumer prices indices (tier) to the FFR (tier 2) and market variables (tier 3). It can be seen that the FFR is impacted by selected production indices that act as rough proxies of broader economic activity. Further, there are interactions within blocks of related variables, e.g., the aggregate industrial production and consumer price indices and their respective constituents. Of particular interest, is the influence exerted by FFR and those variables that influence it with a lag, as seen from Figure~\ref{fig:heatB1}. In particular, FFR impacts consumer price indices positively, which is in accordance with economic theory, which demonstrates its ability in overcoming difficulties in interpretation noted in the literature when small SVAR models were used \citep[see discussion in][]{sims1980macroeconomics,bernanke2005measuring}. Further, it is closely related to the real money stock and treasury yields for both the short and the long tenors, a result in accordance with past analysis \citep{banbura2010large}. 

\subsection{DREAM4 gene expression data}

Next, we briefly discuss how the proposed algorithm can aid in the task of identifying functional relationships (network inference) between genes from limited size gene expression data. This is a fundamental problem in functional genomics and a comprehensive solution to it requires a large set of expensive ``perturbation" (knock-out or knock down) experiments (see discussion in \cite{markowetz2010understand}). The DREAM 4 competition provided data sets to test algorithms for such network inference tasks \citep{marbach2009generating,greenfield2010dream4}, We run the proposed algorithm on a collection of 5 datasets corresponding to different network topologies from selected organisms, each containing time series (21 time points) for 100 genes with 10 perturbations each. Further, to quantify how gains in performance could be achieved in the presence of partial ordering information (which can be obtained from the literature and experimental work in functional genomics applications), we run the proposed algorithm with and without partial ordering. The partial ordering information is constructed by considering the following 3 disjoint sets: ``regulator", ``target" or ``free" that the genes are partitioned into. Specifically, based on a ``gold standard" network of gene functional relationships, that is taken as the underlying truth and denoted by $A$ with $A_{ij}\neq0$ corresponding to an edge $j\rightarrow i$, let
{\small
\begin{equation*}
\begin{split}
    \text{regulator} &:= \big\{i: \forall j,\,A_{ij} = 0~\text{and}~\exists j,\,A_{ji} \neq 0\big\}, \\
    \text{target} &:= \big\{i: \exists j,\,A_{ij} \neq 0~\text{and}~\forall j,\,A_{ji} = 0\big\}; 
    \end{split}
\end{equation*}
}%
that is, the regulator set consists of genes that emit, but have no incoming edges, and the target set contains those that receive, but have no outgoing edges; all other genes including those simultaneously emitting and receiving or neither emitting nor receiving constitute the ``free" set. The partial ordering information is constructed such that we prohibit genes in the regulator set to receive, and those in the target set to emit, while imposing no restrictions on those in the free set. In other words, the partial ordering enforces the regulator and the target sets to form a bipartite graph. 

We run the model on each of the five network topologies datasets. Specifically, for each run, we set $d=1$ and select the hyper-parameters $(\mu_A,\mu_B)$ based on the following procedure: we first run the algorithm over a grid and select the pair $(\mu_A^*,\mu_B^*)$ that gives the smallest predicted RMSE; then we set $\mu_B\equiv\mu_B^*$ and run the algorithm over a sequence of $\mu_A$'s to obtain the ROC/precision-recall curve. Similar to~\cite{lu2021causal}, AUROC and AUPRC are used as performance measures and obtained for each run, with and without the partial ordering information. 

Performance of the proposed algorithm without partial ordering information is in the same ballpark range to other linear methods tested in \cite{lu2021causal}, with AUPRC between 0.10-0.20 and AUROC between 0.60-0.65; for an extensive analysis see tables in~\cite{lu2021causal} and follow-up discussion. Note that many methods exhibiting better performance are non-linear and can accommodate more complex temporal dynamics of gene expression data.  In the presence of partial ordering, we notice a 0.10 increase in AUPRC; the gain in AUROC is of similar magnitude, thus showing the benefits of the proposed method to consume seamlessly such prior information.   

Finally, we note that given the availability of a ``gold standard" for the DREAM4 datasets under consideration, the calculated varsortability averages around 0.40. This suggests that for real world applications, the data scale can be rather uninformative about the underlying topological ordering of the variables,  thus posing challenges for selected continuous structural learning based methods.\footnote{Note that we have effectively ignored temporal dependency while calculating varsortability, and hence the model could potentially be mildly mis-specified. On the other hand, calculation based on simulated data indicates that even though such mis-specification may introduce a minor downward bias to the truth (i.e., the calculated varsortability based on the mis-specified model may underestimate the truth), it will not drastically change the varsortability to a large extent and therefore the conclusion still stands.} This reiterates the need for developing methods for structural discovery in time-series data that are robust to the data scale.

%% file: 06_Conclusion.tex
\section{Conclusion}

The paper presents an efficient algorithm to estimate the parameters of a Structural VAR model, in the presence of a priori information that provides partial ordering information for the variables under consideration. The formulated optimization problem is built upon an existing method that estimates a DAG and augments the objective function with the necessary lag terms that encode the temporal dependency. The acyclicity constraints is enforced through a polynomial number of constraints, which can also seamlessly incorporate the partial ordering information. The proposed algorithm is provably convergent to a stationary point. Numerical experiments on synthetic data illustrate the overall competitive performance of the proposed algorithm to a competing method and also the role of the prior information on the accuracy of the results. Finally, applications to macroeconomic and genomic data demonstrate the usefulness of the algorithm in practical settings.

%% file: A1_admm_update.tex
\section{Implementation Details of the Algorithm}

In this section, we provide additional implementation details of the proposed algorithm, including the exact steps of the proposed ADMM update, as well as notes for practical implementation.

\subsection{Details on the $(A,B,\boldsymbol\lambda)$-update via ADMM}\label{appendix:solveADMM}

In this section, we present the details for cyclically updating each primal block while holding the others fixed, as outlined in Algorithm~2. Note that these updates follow along similar lines to the ones in \citet{yuan2019constrained}, but with necessary modifications to accommodate the additional lag dynamics and partial ordering information that further reduces the search space. Throughout this section, we use $\mathcal{S}_i$ to denote the priori support set corresponding to row $i$ of $A$ and $\widetilde{A}$, based on $\mathcal{P}_A$; in other words, $\mathcal{S}_i:=\{j: A_{ij}\neq 0; A\in\mathcal{P}_A\}$. Note that in the absence of a priori, $\mathcal{S}_i\equiv \{1,\cdots,p\}\backslash\{i\}$. The updating rule for each primal block is given next. Throughout this section, we use $(x)^+ := \max\{0, x\}$ to denote the flooring of $x$ where the positive part is retained otherwise floored at 0.

\paragraph{$A$-update.} Let $\mathbf{V}_n := \mathbf{X}_n - \mathbf{X}_{n-1} B^\top$. For each row $i$, the values of the coordinates in the support set $\mathcal{S}_i$, denoted by $A_{[i,\mathcal{S}_i]}\in\mathbb{R}^{|\mathcal{S}_i|\times 1}$ can be obtained by solving the following linear system:
\begin{equation}\label{eqn:updateArow}
\Big(\frac{1}{n}(\mathbf{X}_n)_{[:,\mathcal{S}_i]}^\top(\mathbf{X}_n)_{[:,\mathcal{S}_i]}+\rho \I_{|\mathcal{S}_{i}|}\Big) A_{[i,\mathcal{S}_i]} = \frac{1}{n}(\mathbf{X}_n)^\top_{[:,\mathcal{S}_i]}(\mathbf{V}_n)_{[:,i]} + \rho\big(\widetilde{A}_{[i,\mathcal{S}_i]}-(U_A)_{[i,\mathcal{S}_i]} \big),
\end{equation}
where $\I_{|\mathcal{S}_i|}$ is an $|\mathcal{S}_i|\times |\mathcal{S}_i|$ identity matrix and $(\mathbf{X}_n)_{[:,\mathcal{S}_i]}$ corresponds to the column subspace indexed by $\mathcal{S}_i$. 
\paragraph{$\widetilde{A}$-update.} The updated $\widetilde{A}$ is given by
\begin{equation*}
\begin{split}
\widetilde{A}  = \argmin\limits_{\widetilde{A}\in\mathcal{P}_A}~&\frac{\rho}{2}\|A - \widetilde{A} + U_A\|_F^2 + \mu_A\sum_{i}\sum_j |\widetilde{A}_{ij}|  \\
& + \frac{\rho}{2}\sum_k\sum_i\sum_{j\neq i}\big(|\widetilde{A}_{ij}|w_{ij} + \tau(1-w_{ij}) + \xi_{ijk} -\tau \lambda_{ik} - \tau \mathbb{I}(j\neq k) + \tau\lambda_{jk} + y_{ijk} \big)^2.
\end{split}
\end{equation*}
The operation is entry-wise. Specifically, for $i,j=1,\cdots,p$, $\widetilde{A}_{ij}\leftarrow 0$ if $j\notin\mathcal{S}_i$; otherwise,
\begin{itemize}[itemsep=0pt,topsep=1pt]
\renewcommand\labelitemi{--}
    \item when $w_{ij}=0$, $\widetilde{A}_{ij}$ is updated by
    \begin{equation*}
    \widetilde{A}_{ij}  \leftarrow \argmin \big\{\frac{\rho}{2} (\widetilde{A}_{ij}-A_{ij}-U_{ij})^2 + \mu_A|\widetilde{A}_{ij}|\big\} = \text{sign}(A_{ij} + U_{ij})\Big(| A_{ij}+ U_{ij} | - \frac{\mu_A}{\rho}\Big)^+;
    \end{equation*}
    \item when $w_{ij}=1$, $\widetilde{A}_{ij}$ is updated by
    \begin{equation*}
    \begin{split}
    \widetilde{A}_{ij} & \leftarrow \argmin \Big\{\frac{\rho}{2}(1+p) \widetilde{A}_{ij}^2 - \rho(A_{ij}+U_{ij}) \widetilde{A}_{ij} + \big(\mu_A+\rho\sum\nolimits_{k}\pi_{ijk}\big)|\widetilde{A}_{ij}|\Big\} \\
    & = \text{sign}(A_{ij} + U_{ij})\Big( \frac{\rho(|A_{ij}+U_{ij}|)-\rho\sum_k\pi_{ijk}-\mu_A}{\rho(1+p)}\Big)^+,
    \end{split}
    \end{equation*}
    where $\pi_{ijk} = \xi_{ijk} - \tau\lambda_{ik} - \tau I(j\neq k) + \tau\lambda_{jk} + y_{ijk}$.
\end{itemize}
\paragraph{$B$-update.} Let $\mathbf{W}_n := \mathbf{X}_n - \mathbf{X}_n A^\top$; for each row $i$ of $B$, denoted by $B_{[i,:]}\in\mathbb{R}^{(dp)\times 1}$, its update can be obtained by solving the following linear system:
\begin{equation*}
    \Big(\frac{1}{n}(\mathbf{X}_{n-1})^\top (\mathbf{X}_{n-1}) + \rho \I_{dp}\Big) B_{[i,:]} = \frac{1}{n}(\mathbf{X}_{n-1})^\top(\mathbf{W}_n)_{[:,i]} + \rho\big(\widetilde{B}_{[i,:]} - (U_B)_{[i,:]}\big).
\end{equation*}
\paragraph{$\widetilde{B}$-update.} The update of $\widetilde{B}$ can be obtained by entry-wise soft-thresholding:
\begin{equation*}
    \widetilde{B}_{ij} \leftarrow \text{sign}\big(B_{ij}+(U_B)_{ij}\big) \Big(|B_{ij}+(U_B)_{ij}|-\frac{\mu_B}{\rho}\Big)^{+}.
\end{equation*}
\paragraph{$\boldsymbol{\lambda}$-update.} The update of $\boldsymbol{\lambda}$ satisfies
\begin{equation}\label{eqn:lambda0}
\boldsymbol{\lambda}^\star := \argmin_{\boldsymbol{\lambda}} f(\boldsymbol{\lambda}) = \sum_{k} \sum_{i \neq j} \left(|\widetilde{A}_{ij}|w_{ij} + \tau(1-w_{ij})+\xi_{ijk}- \tau\lambda_{ik} - \tau \mathbb{I}(j\neq k)+\tau\lambda_{jk} + y_{ijk} \right)^2.
\end{equation}
Here for stability consideration, we consider instead
\begin{equation}\label{eqn:lambda}
\min\limits_{\boldsymbol{\lambda}}\quad \widetilde{f}_{\phi}(\boldsymbol{\lambda}) :=  f(\boldsymbol{\lambda}) + \phi\cdot\mathrm{tr}(\boldsymbol{\lambda}^\top \boldsymbol{\lambda}), \qquad \text{for some}~~\phi >0.
\end{equation}
Denote the solution to~\eqref{eqn:lambda} by $\widetilde{\boldsymbol{\lambda}}(\phi)$ which is a function of $\phi$; the desired optimizer $\boldsymbol{\lambda}^*$ to~\eqref{eqn:lambda0} then corresponds to $\boldsymbol{\lambda}^* = \lim\limits_{\phi\rightarrow 0^+}\widetilde{\boldsymbol{\lambda}}(\phi)$. After some algebra, this gives rise to the following update rule:
\begin{equation*}
\boldsymbol{\lambda} \leftarrow -\frac{1}{2\tau p}\Big[\I_p + \frac{1}{2p} \mathbf{1}\mathbf{1}^\top \Big]\psi,
\end{equation*}
where $\mathbf{1} := (1,\cdots,1)^\top \in\mathbb{R}^{p\times 1}$, $\boldsymbol{\psi}\in\mathbb{R}^{p\times p}$ collects the terms associated with $\widetilde{A},\boldsymbol{w},\boldsymbol{\xi}$ and $y$:
\begin{equation*}
\begin{split}
\psi_{ik} & = \sum_{j\neq i} \big(|\widetilde{A}_{ji}|w_{ji}-|\widetilde{A}_{ij}|w_{ij}\big) - \tau\sum_{j\neq i}(w_{ji}-w_{ij}) + \sum_{j\neq i}\big(\xi_{jik}-\xi_{ijk}\big)+\sum_{j\neq i}\big(y_{jik}-y_{ijk}\big) \\
&\quad -\tau\big(\sum_{j\neq i}\mathbb{I}(i\neq k) -\sum_{j\neq i}\mathbb{I}(j\neq k)\big).
\end{split}
\end{equation*}

\paragraph{$\boldsymbol{\xi}$-update.}
\begin{equation*}
\xi_{ijk} = \max\Big\{0, \tau\lambda_{ik} + \tau\mathbb{I}(j\neq k) - \tau \lambda_{jk} - |\widetilde{A}_{ij}|w_{ij} - y_{ijk} - \tau(1-w_{ij}) \Big\}.
\end{equation*}

\subsection{Initialization and the choice of hyper-parameters}\label{appendix:hparams}

\paragraph{Initialization.} In our implementation, we use the following initialization: $\boldsymbol{w}=\mathbf{1}, A^{(0)}=\mathbf{0},B^{(0)}=\mathbf{0}$. One can consider an alternative initialization with
\begin{equation*}
    B^{(0)} = \argmin\nolimits_{B} \ell(A^{(0)}, B; \mathbf{X}_n, \mathbf{X}_{n-1}) + \mu_B\|B\|_1,
\end{equation*}
which in practice, makes little difference to the result. 

\paragraph{The choice of hyper-parameters.} There are two types of ``hyper-parameters" involved in the optimization: (1) $\mu_A,\mu_B$ that enter the regularization term and govern the sparsity of the estimated $A,B$'s, and (2) ``algorithmic" ones such as $\tau, \rho$ and the convergence tolerance that determines the termination of the updates. 

For hyper-parameters of the first type, that is, $(\mu_A,\mu_B)$, we advise to choose them over a lattice, with the best combination determined by the out-of-sample root-mean-squared-error (RMSE). More concretely, given observed time series data $\{x_1,\cdots,x_n\}$, after constructing them as (target, predictors) pairs, given by $(x_t, \{x_t, x_{t-1},\cdots,x_{t-q}\}),t=q+1,...,n$, one can split the data into train-validation sets; after obtaining the model parameters based on the train sets, RMSE is calculated on the validation set and the optimal combination of the hyper-parameters are determined by the one that minimizes the validation RMSE on the lattice\footnote{One can alternatively create $K$ disjoint folds of the samples and obtain the out-of-sample RMSE based on $K$-fold cross-validation.} Note that the temporal dependency between target and predictors is retained given that the train-validation spilt are created based on the pairs. Practically, the search is first conducted on a coaser grid that encompasses a wide range, then refined on a finer one. For the settings considered in this paper, $\mu_A$ is initially chosen from the range $[0.01, 0.3]$ and $\mu_B$ is chosen as $c\sqrt{\log(pq)/N}$, where $c$ ranges from $[0.05,0.5]$; $p$ is the number of nodes, $q$ the number of lags and $N$ the effective number of sample size $N:=n-q$. 

For ``algorithmic" hyper-parameters $\tau$ and $\rho$ that respectively show up in the $\ell_1$ truncated norm and the augmented Lagrangian, we notice that the performance is reasonably robust to how they are chosen. In our experiments, we fix them respectively at $\tau=10^{-6}$ and $\rho=1$. Regarding the convergence criterion for the ADMM update, we advise to set it commensurately to the available sample size; it is empirically observed that in sample-deprived settings, a tighter convergence tolerance can typically yield better performance, at the compromise of more iterations and hence longer computing time, whereas in setting where sample size is moderate, a more relaxed convergence tolerance can perform reasonably well. 

Finally, note that in this work, we have fixed the number of lags at 2, in accordance with the underlying truth in synthetic data experiments. In practice, there are several possible routes that one can adopt:  (1) search it in conjunction with $(\mu_A,\mu_B)$ by expanding the lattice to a 3D grid; or (2) by examine the magnitude of the estimation coefficients and determine the number of lags together with some prior knowledge/existing literature on the subject matter. 

%% file: A2_algo_convergence.tex
\section{Algorithmic Convergence}\label{appendix:algo_conv}

We start by providing the necessary definitions used in the proof of the main result (Proposition~1).

\begin{definition}
For any subset $\mathcal{S} \subseteq \mathbb{R}^d$ and any point 
$\boldsymbol{x}\in\mathbb{R}^d$, the distance from
$\boldsymbol{x}$ to $\mathcal{S}$ is defined and denoted by
$\text{dist}(\boldsymbol{s}, \mathcal{S}) := \inf \{ \|\boldsymbol{y}- \boldsymbol{x}\|: \boldsymbol{y} \in \mathcal{S}\}$.
When $\mathcal{S} = \emptyset$, then $\text{dist}(\boldsymbol{x}, \mathcal{S}) = \infty$ for all $\boldsymbol{x}$.
\end{definition}

\begin{definition}
Let $c \in [0,\infty)$. Denote by $\Gamma_{c}$ the class of all functions $\phi: [0,c]\rightarrow \mathbb{R}^+$ that satisfy the following conditions:
\begin{enumerate}[itemsep=0pt,topsep=1pt]
  \item $\phi$ is continuous on $[0,c)$;
  \item $\phi$ is concave on $(0,c)$;
  \item $\phi(0)=0$, $\phi'(s)> 0, ~\forall~s \in (0,c)$, where $\phi'(s)$ is the derivative of $\phi$ evaluated at $s$.
\end{enumerate}
\end{definition}
  
\begin{definition}[Fr\'{e}chet subdifferential $\widehat{\partial} f$ and limiting subdifferential $\partial f$] Denote the domain of a function $f: \mathbb{R}^d\mapsto \mathbb{R}$ by $\text{dom}(f) := \{\boldsymbol{x} \in\mathbb{R}^d:f(x)<\infty\}$.
\begin{enumerate}[itemsep=0pt,topsep=1pt]
    \item The Fr\'{e}chet subdifferential of $f$ at $\boldsymbol{x}$ is defined as
\begin{equation}\label{eq:frechet}
\widehat{\partial}f(\boldsymbol{x})=\{\boldsymbol{x}^\star \in\mathbb{R}^d:\liminf_{\boldsymbol{y}\neq \boldsymbol{x}: \boldsymbol{y}\rightarrow \boldsymbol{x}} \frac{f(\boldsymbol{y})-f(\boldsymbol{x})- \langle \boldsymbol{x}^\star, \boldsymbol{y}- \boldsymbol{x}\rangle}{\|\boldsymbol{y}-\boldsymbol{x}\|}
\geq 0\}
\end{equation}
\item The limiting subdifferential \citep{mordukhovich2006variational} of $f$ at $\boldsymbol{x} \in\text{dom}(f)$, denoted by $\partial f(\boldsymbol{x})$, is defined as follows:
\begin{equation}\label{eq:subdifferential}
\partial f(\boldsymbol{x}) := \{\boldsymbol{x}^\star \in\mathbb{R}^d: 
\exists~\boldsymbol{x}^k \rightarrow \boldsymbol{x}, ~f(\boldsymbol{x}^k) \rightarrow f(\boldsymbol{x})~\text{and}~\boldsymbol{x}^k \in \widehat{\partial}f (\boldsymbol{x}^k) \rightarrow \boldsymbol{x}^\star~\text{as}~k\rightarrow\infty\}.
\end{equation}
\end{enumerate}
\end{definition}

\begin{definition}[Kurdyka-\L{}ojasiewicz property]
A function $f:\mathbb{R}^d\rightarrow (-\infty,+\infty]$ possesses the Kurdyka-\L{}ojasiewicz (K-L) 
property at point $\boldsymbol{z}_0 \in \text{dom}(\partial f)$, if there exist $c>0$, a neighborhood $\mathcal{Z}$ of $\boldsymbol{z}_0$ and a function $\phi \in \Gamma_{c}$, such that for all 
$$\{\boldsymbol{z}\in \mathcal{Z}\} \bigcap \big\{\boldsymbol{z}: f(\boldsymbol{z}_0) <f(\boldsymbol{z})<f(\boldsymbol{z}_0)+ c\big\},$$ the following inequality holds
\begin{equation*}
 \phi'\big(f(\boldsymbol{z})-f(\boldsymbol{z}_0)\big) \text{dist} \big(0,\partial f(\boldsymbol{z})\big)\geq 1.
\end{equation*}
\end{definition}

\begin{definition}[Semi-algebraic sets and functions]\hfill
\begin{enumerate}[itemsep=0pt,topsep=1pt]
\setlength\itemsep{0pt}
\item A subset $\mathcal{C} \in \mathbb{R}^{d\times d}$ is semi-algebraic, if there exists a finite number of real polynomial functions $\{g_{ij}\}, \{s_{ij}\}: \mathbb{R}^{d \times d} \mapsto \mathbb{R}$ such that 
$$
\mathcal{C} =\bigcup\nolimits_{i=1}^{\bar{p}}\bigcap\nolimits_{j=1}^{\bar{q}}\big\{\boldsymbol{z}\in \mathbb{R}^{d \times d}:~g_{ij} (\boldsymbol{z})=0 ~~\text{and} ~~s_{ij}(\boldsymbol{z})<0\big\}.
$$
\item A function $h: \mathbb{R}^{d \times d}\rightarrow (-\infty, +\infty]$ is called semi-algebraic, if its graph
$$
\mathcal{G}(h):= \big\{(\boldsymbol{z},y)\in \mathbb{R}^{(d\times d)+1}: h(\boldsymbol{z})=y\big\},
$$
is a semi-algebraic set in $\mathbb{R}^{(d \times d)+1}$.
\end{enumerate}
\end{definition}

\begin{definition}[Sub-analytic sets and functions]\hfill
\begin{enumerate}[itemsep=0pt,topsep=1pt]
\setlength\itemsep{0pt}
\item A subset $\mathcal{C}\in \mathbb{R}^{d \times d}$ is sub-analytic, if there exists a finite number of real analytic functions $\{g_{ij}\},\{s_{ij}\}: \mathbb{R}^{d \times d} \rightarrow \mathbb{R}$ such that 
$$
\mathcal{C}=\bigcup\nolimits_{i=1}^{\bar{p}}\bigcap\nolimits_{j=1}^{\bar{q}}\big\{ \boldsymbol{z}\in \mathbb{R}^d: g_{ij}(\boldsymbol{z})=0 ~~ \text{and} ~~s_{ij}(\boldsymbol{z})<0\big\}.
$$
\item A function $h$: $\mathbb{R}^{d \times d}\rightarrow (-\infty, +\infty]$ is called sub-analytic, if its graph
$$
\mathcal{G}(h):= \big\{(\boldsymbol{z},y)\in \mathbb{R}^{(d \times d)+1}: h(\boldsymbol{z}) = y\big\}
$$
is a sub-analytic set in $\mathbb{R}^{(d \times d)+1}$.
\end{enumerate}
\end{definition}

\noindent
\begin{remark}\label{remark-KL}
As noted in \cite{bolte2014proximal}, both real analytic and semi-algebraic functions are sub-analytic. In general, the sum of two sub-analytic functions is not necessarily sub-analytic. However, it can be shown that for two sub-analytic functions, if at least one function maps bounded sets to bounded sets, then their sum is also sub-analytic (see discussion and examples in \cite{bolte2014proximal}).
\end{remark}

\begin{lemma}\label{lemma-KL}
The augmented Lagrangian function in Algorithm~2
\begin{equation}\label{eq:augm-lagrangian}
\begin{split}
L_\rho(A,\widetilde{A},&B,\widetilde{B},\boldsymbol{\lambda},\boldsymbol{\xi}; U_A,U_B,\boldsymbol{y}) 
=~\ell(A,B; \mathbf{X}_n,\mathbf{X}_{n-1}) + \mu_A\|\widetilde{A}\|_1 + \mu_B \|\widetilde{B}\|_1\\
& + \frac{\rho}{2}\|A - \widetilde{A}\|_{\F}^2  + \rho\langle A-\widetilde{A}, U_A\rangle + \frac{\rho}{2}\|B - \widetilde{B}\|_{\F}^2  + \rho\langle B-\widetilde{B}, U_B\rangle\\
& + \frac{\rho}{2}\sum_k\sum_{i \neq j}\big( |\widetilde{A}_{ij}|w_{ij} + \tau(1-w_{ij}) + \xi_{ijk} - \tau \lambda_{ik} - \tau \mathbb{I}(j\neq k) + \tau\lambda_{jk}\big)^2  \\
& + \rho\sum_k\sum_{i \neq j}y_{ijk}\big( |\widetilde{A}_{ij}|w_{ij} + \tau(1-w_{ij}) + \xi_{ijk} - \tau \lambda_{ik} - \tau \mathbb{I}(j\neq k) + \tau\lambda_{jk}\big)
 \end{split}
\end{equation}
satisfies the K-L property.
\end{lemma}

\begin{proof}
We consider each term in \eqref{eq:augm-lagrangian}.
The loss function and the two terms involving the Frobenius norm are analytic functions. Further, the two
$\ell_1$ penalty terms are sub-algebraic (see Example 3, Section 5 in \cite{bolte2014proximal}). Finally, the remaining terms in \eqref{eq:augm-lagrangian} are also sub-algebraic (see Example 2, Section 5 in \cite{bolte2014proximal}). Further, the sub-algebraic functions in \eqref{eq:augm-lagrangian} map bounded sets to bounded sets. Hence, by Remark \ref{remark-KL}, it follows that the augmented Lagrangian $L_\rho(\cdot)$ satisfies the K-L property.
\end{proof}

Let $\Theta=(A,\widetilde{A},B,\widetilde{B},\boldsymbol{\lambda},\boldsymbol{\xi})$ and $\Psi=(U_A,U_B,\boldsymbol{y})$ denote the collection of primal and dual variables, respectively.

\begin{lemma}\label{lemma-sandwich}
Let $(\Theta^{(s)},\Psi^{(s)})$ be a sequence generated by Algorithm~2. Then, there exists $\eta>0$ such that
{\small
\begin{align}
L_\rho&(\Theta^{(s+1)};\Psi^{(s+1)})
    \leq  L_\rho(\Theta^{s};\Psi^{s})
    -\frac{\eta}{2}\Big[\|A^{(s)}-A^{(s+1)}\|_\F^2 + \|\widetilde{A}^{(s)}-\widetilde{A}^{(s+1)}\|_\F^2 \|B^{(s)}-B^{(s+1)}\|_\F^2 + \|\widetilde{B}^{(s)}-\widetilde{B}^{(s+1)}\|_\F^2  \nonumber \\ 
     & +\|\boldsymbol{\lambda}^{(s)}-\boldsymbol{\lambda}^{(s+1)}\|_\F^2 + \|\boldsymbol\xi^{(s)}-\boldsymbol\xi^{(s+1)}\|_\F^2
    + \|U_A^{(s)}-U_A^{(s+1)}\|_\F^2 + \|U_B^{(s)}-U_B^{(s+1)}\|_\F^2 + \|\boldsymbol y^{(s)}-\boldsymbol y^{(s+1)}\|_\F^2 
    \Big].\label{eq:lemma-1}
\end{align}
}%
\end{lemma}

\begin{proof}
The proof proceeds by considering updates of every primal and dual variable of the augmented Lagrangian function. We start by considering the update of primal variable $A$. Note that the loss function $\ell(\cdot,\cdot;~\text{data})$ is convex in $A$, for any fixed $B$. Hence, using the first order optimality conditions for $A$ and the convexity of $\ell$ in $A$, we get
\begin{eqnarray}\label{eq:lemma-1-A}
0 & = & \big\langle A^{(s)}-A^{(s+1)},~ \nabla_A\ell(A^{(s+1)},B^{(s)};\mathbf{X}_n,\mathbf{X}_{n-1}) +\rho(A^{(s+1)}-\widetilde{A}^{(s)})+\rho U_A^{(s)} \big\rangle \\ \nonumber &\leq &\ell(A^{(s)},B^{(s)};\mathbf{X}_n,\mathbf{X}_{n-1})-\ell(A^{(s+1)},B^{(s)};\mathbf{X}_n,\mathbf{X}_{n-1}) \\
& & ~+ \rho\big\langle A^{(s+1)}-\widetilde{A}^{(s)}, A^{(s)}-A^{(s+1)}\big\rangle +\rho\langle U_A^{(s)},A^{(s)}-A^{(s+1)}\big\rangle \\ \nonumber
& = &\ell(A^{(s)},B^{(s)};\mathbf{X}_n,\mathbf{X}_{n-1})-\ell(A^{(s+1)},B^{(s)};\mathbf{X}_n,\mathbf{X}_{n-1}) +  \rho \big\langle U_A^{(s)}, A^{(s)}-A^{(s+1)}\big\rangle \\ \nonumber
& &~
- \frac{\rho}{2} \|A^{(s)}-A^{(s+1)}\|_\F^2 +\frac{\rho}{2} \|A^{(s)}-\widetilde{A}^{(s)}\|_\F^2 -\frac{\rho}{2}\|A^{(s+1)}-\widetilde{A}^{(s)}\|_\F^2 \\ \nonumber
& = & L_\rho(\Theta^{(s)},\Psi^{(s)})-L_\rho(A^{(s+1)},\widetilde{A}^{(s)},B^{(s)},\widetilde{B}^{(s)},\boldsymbol{\lambda}^{(s)},\boldsymbol{\xi}^{(s)};\Psi^{(s)})-\frac{\rho}{2}\|A^{(s)}-A^{(s+1)}\|_\F^2, \nonumber
\end{eqnarray}
where the first inequality follows from the first order characterization of the convex loss function $\ell$, the second equality using the fact 
  $$(u_1-u_2)^\top(u_3-u_1)=\frac{1}{2}\Big( \|u_2-u_3\|^2_{\F}-\|u_1-u_2\|^2_{\F}-\|u_1-u_3\|^2_{\F}\Big),$$
and the third equality adding and subtracting the additional terms in the augmented Lagrangian function to obtain the corresponding expressions for $L_\rho$. 

Next, we consider the update for primal variable $\widetilde{A}$. Then, using Lemma \ref{lemma-suff-dec} together with adding and subtracting the additional terms in the augmented Lagrangian function, we obtain
\begin{equation}\label{eq:lemma-1-tildeA}
0\leq L_\rho(\Theta^{(s)},\Psi^{(s)})- L_\rho(A^{(s+1)},\widetilde{A}^{(s+1)},B^{(s)},\widetilde{B}^{(s)},\boldsymbol{\lambda}^{(s)},\boldsymbol{\xi}^{(s)};\Psi^{(s)})-\frac{\mu_A-\zeta_{h_1}}{2}\|\widetilde{A}^{(s)}-\widetilde{A}^{(s+1)}\|_\F^2,
\end{equation}
where $\zeta_{h_1}$ is the Lipschitz constant in the function $h_1(\widetilde{A})$ (refer to the statement of Lemma \ref{lemma-suff-dec} for the definition of $h_1(\cdot)$).

Analogous derivations to the updates for $A$ and $\widetilde{A}$ hold for the updates of $B$ and $\widetilde{B}$, respectively.

Next, we consider the update for primal variable $\boldsymbol{\lambda}$. Note that it is obtained by solving the problem posited in \eqref{eqn:lambda}. The latter comprises of a quadratic function in $\boldsymbol{\lambda}$ which is continuously differentiable, and a proper lower semi-continuous function in $\boldsymbol{\lambda}$ with $\inf_{\boldsymbol{\lambda}} f(\boldsymbol{\lambda})>-\infty$. Hence,
an application of Lemma \ref{lemma-suff-dec} applied to $\widetilde{f}_\phi(\boldsymbol{\lambda})$ establishes 
\begin{equation}\label{eq:lemma-1-lambda}
0\leq L_\rho(\Theta^{(s)},\Psi^{(s)}) -  L_\rho(A^{(s+1)},\widetilde{A}^{(s+1)},B^{(s+1)},\widetilde{B}^{(s+1)},\boldsymbol\lambda^{(s+1)},\boldsymbol\xi^{(s)};\Psi^{(s)})-\frac{\phi-1}{2}\|\boldsymbol\lambda^{(s)}-\boldsymbol\lambda^{(s+1)}\|_\F^2.
\end{equation}

The update for primal variable $\boldsymbol\xi$ solves an objective function comprising of a quadratic function and a proper lower semi-continuous function in $\boldsymbol\xi$, so a similar derivation as for $\boldsymbol\lambda$ establishes the result.

Next, we consider the update for dual variable $U_A$. Note that it admits an explicit update $U_A^{(s+1)}=U_A^{(s)}+(A^{(s)}-\widetilde{A}^{(s+1)})$. Further, note that $L_\rho(\Theta^{(s)};\Psi^{(s)})-L_\rho(\Theta^{(s)}; U_A^{(s+1)},U_B^{(s)},\boldsymbol y^{(s)})=\rho\langle A^{(s+1)}-\widetilde{A}^{(s+1)}, U_A^{(s)}-U_A^{(s+1)}\rangle=
\rho\|U_A^{(s)}-U_A^{(s+1)}\|_\F^2$, where the second equality follows from the exact update for $U_A$. Hence, we obtain
\begin{equation*}
0\leq L_\rho(\Theta^{(s)},\Psi^{(s)}) -  L_\rho(\Theta^{(s+1)};U_A^{(s+1)},U_B^{(s)},\boldsymbol y^{(s)})-\frac{\rho}{2}\|U_A^{(s)}-U_A^{(s+1)}\|_\F^2.
\end{equation*}

An analogous derivation to that for $U_A$ holds for dual
variables $U_B$ and $\boldsymbol y$.

Then, selecting $\eta=\max\{\rho, \rho\mu_A-\zeta_{h_1},\rho\phi-1\}$, we obtain the desired result.
\end{proof}

The following Lemma establishes the sufficient decrease property of the augmented Lagrangian function, after a proximal map step computed for the update of $\widetilde{A}$ (see Section~\ref{appendix:solveADMM} in the supplementary material for the exact update formulas).

\begin{lemma}[Sufficient decrease property for $\widetilde{A}$]\label{lemma-suff-dec}
Let $h_1(\widetilde{A}) :=\|A-\widetilde{A}+U\|_\F^2$, which is a continuously differentiable function with gradient $\nabla{h}_1$ being Lipschitz continuous with Lipschitz constant $\zeta_{h_{1}}$. 
Further, let $h_2(\widetilde{A}) :=\mu_A \|\widetilde{A}\|_1+\frac{\rho}{2}\sum\nolimits_k\sum\nolimits_{i \neq j}\big( |\widetilde{A}_{ij}|w_{ij} + \tau(1-w_{ij}) + \xi_{ijk} - \tau \lambda_{ik} - \tau \mathbb{I}(j\neq k) + \tau\lambda_{jk}\big)^2$, which is 
a proper lower semi-continuous function with $\inf_{\widetilde{A}} h_2(\widetilde{A}) >-\infty$. Let $\mu_A>\zeta_{h_1}$, Then, we have
\begin{eqnarray*}
  h_1\big(\widetilde{A}^{(s+1)}\big)+ h_2\big(\widetilde{A}^{(s+1)}\big) &\leq & h_1\big(\widetilde{A}^{(s)}\big)+ h_2\big(\widetilde{A}^{(s)}\big) - \frac{(\mu_A - \zeta_{h_1})}{2}\|\widetilde{A}^{(s+1)}-\widetilde{A}^{(s)}\|_\F^2.
\end{eqnarray*}

\end{lemma}

\begin{proof}
Note that due to the presence of the regularizer in $h_2$, the update for $\widetilde{A}$ corresponds to a proximal map step. By the definition of the proximal map, that is, 
\begin{equation*}
\text{prox}(h_2(\widetilde{A});\mu_A):=\argmin_{C\in\mathbb{R}^{p\times p}} \Big\{h_2(C) + \frac{\mu_A}{2} \| C-\widetilde{A}\|^2_{\F}\Big\},
\end{equation*}
a first-order Taylor expansion around $A^{(s)}$ gives
\begin{equation*}
\widetilde{A}^{(s+1)}= \argmin_{\widetilde{A}\in\mathbb{R}^{p\times p}} \bigl\{
\langle \widetilde{A} -\widetilde{A}^{(s)}, \nabla h_1(\widetilde{A}^{(s)}) \rangle +\frac{\mu_A}{2} \|\widetilde{A}-\widetilde{A}^{(s)}\|^2_{\F} + h_2(\widetilde{A}) \bigr\}.
\end{equation*}
Thus, by setting $\widetilde{A} \equiv \widetilde{A}^{(s+1)}$,  we obtain the following based on the optimality of $\widetilde{A}^{(s+1)}$:
\begin{equation}\label{eq:proximal-map-bound}
\langle \widetilde{A}^{(s+1)}-\widetilde{A}^{(s)}, \nabla h_1(\widetilde{A}^{(s)}) \rangle
+\frac{\mu_A}{2}\|\widetilde{A}^{(s+1)}-\widetilde{A}^{(s)}\|_{\F}^2+
h_2\big(\widetilde{A}^{(s+1)}\big) \leq h_2\big(\widetilde{A}^{(s)}\big).
\end{equation}
Using first the Lipschitz continuity of the gradient of $h_1$ and then leveraging \eqref{eq:proximal-map-bound}, we get
\begin{align*}
h_1(\widetilde{A}^{(s+1)})+h_2(\widetilde{A}^{(s+1)}) & \leq 
h_1(\widetilde{A}^{(s)}) + \langle \widetilde{A}^{(s+1)}-\widetilde{A}^{(s)},\nabla h_1(\widetilde{A}^{(s)}\rangle +\frac{\zeta_{h_1}}{2} \|\widetilde{A}^{(s)}-\widetilde{A}^{(s+1)}\|_\F^2 + h_2(\widetilde{A}^{(s+1)}) \\ 
&  \leq h_1(\widetilde{A}^{(s)}+h_2(\widetilde{A}^{(s)}) - \frac{\mu_A-\zeta_{h_1}}{2}\|\widetilde{A}^{(s+1)}-\widetilde{A}^{(s)}\|_{\F}^2.
\end{align*}

The condition $\mu_A>\zeta_{h_1}$ ensures the sufficient decrease property in $h_1(\widetilde{A}^{(s+1)})+ h_2(\widetilde{A}^{(s+1)})$.
\end{proof}

\begin{lemma}\label{lemma-subsequence}
Let $(\Theta^{(s)},\Psi^{(s)})$ be a sequence generated by Algorithm~2. Then, there exists a subsequence $(\Theta^{(s_v)},\Psi^{(s_v)})$ of $(\Theta^{(s)},\Psi^{(s)})$, such that
\begin{equation}\label{eq:limits-1}
    \lim_{v\rightarrow\infty} \ell(A^{(s_v)},B^{(s_v)}; \mathbf{X}_n,\mathbf{X}_{n-1})=\ell(A^*,B^*;\mathbf{X}_n,\mathbf{X}_{n-1}),
\end{equation}
\begin{equation}\label{eq:limits-2}
\lim_{v\rightarrow \infty} \mu_A \|\widetilde{A}^{(s_v)}\|_1 = \mu_A\|A^*\|_1, \ \
    \lim_{v\rightarrow \infty} \mu_B\|\widetilde{B}^{(s_v)}\|_1 = \mu_B\|B^*\|_1, 
\end{equation}
\begin{align}\label{eq:limits-3}
    \lim_{s_v\rightarrow\infty} \sum_k\sum_{i \neq j}\big( |\widetilde{A}^{(s_v)}_{ij}|w_{ij} + \tau(1-w_{ij}) + \xi_{ijk} - \tau \lambda^{(s_v)}_{ik} - \tau \mathbb{I}(j\neq k) + \tau\lambda^{(s_v)}_{jk}\big)^2 = \\ \nonumber
    \sum_k\sum_{i \neq j}\big( |\widetilde{A}^*_{ij}|w_{ij} + \tau(1-w_{ij}) + \xi_{ijk} - \tau \lambda^*_{ik} - \tau \mathbb{I}(j\neq k) + \tau\lambda^*_{jk}\big)^2  
\end{align}
where 
$$(\Theta^*,\Psi^*) := \lim_{v\rightarrow \infty} (\Theta^{(s_v)},\Psi^{(s_v)}).$$
\end{lemma}

\begin{proof}
We consider the constituent parts of the augmented Lagrangian function. The loss function  $\ell(\cdot,\cdot~;\text{data})$ and the regularization terms ($\ell_1$ norms) are bounded below. 

Next, consider the term $\frac{\rho}{2}\|B-\widetilde B\|^2_{\F} + \rho\langle B-\widetilde B,U_B \rangle$. Some calculations show that it is a positive semi-definite function in its arguments $(B,\widetilde B,U_B)$ and hence coercive and therefore bounded below. A similar calculation shows that the term $\frac{\rho}{2}\|A-\widetilde A\|^2_{\F} + \rho \langle A-\widetilde A,U_A \rangle+ \frac{\rho}{2}\sum\nolimits_k\sum\nolimits_{i \neq j}\big( |\widetilde{A}_{ij}|w_{ij} + \tau(1-w_{ij}) + \xi_{ijk} - \tau \lambda_{ik} - \tau \mathbb{I}(j\neq k) + \tau\lambda_{jk}\big)^2 
 + \rho\sum\nolimits_k\sum\nolimits_{i \neq j}y_{ijk}\big( |\widetilde{A}_{ij}|w_{ij} + \tau(1-w_{ij}) + \xi_{ijk} - \tau \lambda_{ik} - \tau \mathbb{I}(j\neq k) + \tau\lambda_{jk}\big),
$ is also a positive semi-definite function in its arguments $(A,\widetilde{A},U_A,\boldsymbol{\lambda},\boldsymbol{\xi},\boldsymbol{y})$ and thus coercive. Since the sum of coercive functions is also coercive, while the sum of a coercive function and a continuous function bounded below is also coercive, we get that the augmented Lagrangian function is coercive and bounded below. Hence, there exists a function $\underbar{L}(A,\widetilde{A},B,\widetilde{B},\boldsymbol{\lambda},\boldsymbol{\xi};U_A,U_B,\boldsymbol{y})$ such that
\begin{equation}\label{eq:lower-bound-Lagrangian}
L_\rho(\Theta^{(s)},\Psi^{(s)})\geq \underbar{L}(\Theta,\Psi),~\forall\Theta,\Psi.
\end{equation}
Next, using the result in Lemma \ref{lemma-sandwich} we obtain, for $S\in\mathbb{Z}^+$ and $\forall \Phi,\Psi$, 
{\small
\begin{align}\label{eq:bounded-sequence}
& \frac{\eta}{2}\sum_{s=0}^S \bigg[\|A^{(s)}-A^{(s+1)}\|_\F^2 + \|\widetilde{A}^{(s)}-\widetilde{A}^{(s+1)}\|_\F^2 + \|B^{(s)}-B^{(s+1)}\|_\F^2 + \|\widetilde{B}^{(s)}-\widetilde{B}^{(s+1)}\|_\F^2    +\|\boldsymbol{\lambda}^{(s)}-\boldsymbol{\lambda}^{(s+1)}\|_\F^2 \nonumber \\
& + \|\boldsymbol\xi^{(s)}-\boldsymbol\xi^{(s+1)}\|_\F^2
+ \|U_A^{(s)}-U_A^{(s+1)}\|_\F^2 + \|U_B^{(s)}-U_B^{(s+1)}\|_\F^2 + \|\boldsymbol y^{(s)}-\boldsymbol y^{(s+1)}\|_\F^2 
\bigg]\leq L_\rho(\Theta^0,\Psi^0) - \underbar{L}(\Theta,\Psi),
\end{align}
}%
which establishes that $L_\rho(\Theta^{(s)},\Psi^{(s)})$ converges to $L_\rho(\Theta^*,\Psi^*)$. Then, \eqref{eq:bounded-sequence} together with the coerciveness of the augmented Lagrangian functions implies that the entire sequence $(\Theta^{(s)},\Psi^{(s)})$ is a bounded one. The latter fact implies that there exists a subsequence %
\begin{equation*}
(\Theta^{(s_v)},\Psi^{(s_v)}):=(A^{(s_v)},\widetilde{A}^{(s_v)},B^{(s_v)},\widetilde{B}^{(s_v)},\boldsymbol{\lambda}^{(s_v)},\boldsymbol{\xi}^{(s_v)}; U_A^{(s_v)},U_B^{(s_v)},\boldsymbol{y}^{(s_v)}), \ \ v=0,1,\cdots,
\end{equation*}
such that $\lim_{v\rightarrow\infty} (\Theta^{(s_v)},\Psi^{(s_v)})=(\Theta^*,\Psi^*)$.

Next, using the facts that $\ell(A,B;\text{data})$, the regularizer terms, and $\sum_k\sum_{i \neq j}\big( |\widetilde{A}^{(s_v)}_{ij}|w_{ij} + \tau(1-w_{ij}) + \xi_{ijk} - \tau \lambda^{(s_v)}_{ik} - \tau \mathbb{I}(j\neq k) + \tau\lambda^{(s_v)}_{jk}\big)^2$ are continuous functions in their respective arguments establish \eqref{eq:limits-1}, \eqref{eq:limits-2} and \eqref{eq:limits-3}.
\end{proof}

\begin{lemma}\label{lemma-limit-point}
Algorithm~2 either stops at a stationary point of $L_\rho(\Theta,\Psi)$, or generates an infinite sequence such that any limit point of $\{(\Theta^{(s)},\Psi^{(s)})\}$ is a critical point of $L_\rho(\Theta^{(s)},\Psi^{(s)})$.
\end{lemma}

\begin{proof}
To prove this lemma, we leverage the surrogate derivation presented next. 

\paragraph{A surrogate derivation.} Consider the following optimization problem
\begin{equation}\label{eq:general-problem}
\min_{x_1,\cdots,x_J} \sum_{j=1}^J f_j(x_j) \ \ \text{subject to} \ \sum_{j=1}^J C_j x_j=b, \ \ x_j\in\mathcal{X}_j,
\end{equation}
where $\mathcal{X}_j\subseteq \mathbb{R}^{m_j}, C_j\in\mathbb{R}^{p\times m_j}, b\in\mathbb{R}^p$, and $f_j: \mathbb{R}^{m_j}\rightarrow \mathbb{R}$ are closed, convex functions.

The augmented Langrangian function is given by
\begin{equation}\label{eq:general-augmented-L}
L_\rho(x_1,\cdots,x_J;\Lambda)=\sum_{j=J}^J f_j(x_j)-\langle \Lambda,\sum_{j=1}^J A_jx_j-b\rangle +\frac{\rho}{2}\| \sum_{j=1}^J A_j x_j-b\|_2^2,
\end{equation}
with $\Lambda^\top\in\mathbb{R}^p$ the dual variable (Lagrange multiplier) and $\rho>0$ a scaling parameter.

Then, the updates of a multi-block ADMM algorithm to solve \eqref{eq:general-augmented-L} are given by:
\begin{align}
x_j^{(s+1)} & =\argmin_{x_j\in\mathcal{X}_j} \nabla f_j(x_j)+
\frac{\rho}{2}
\|\sum_{k=1}^j A_k x_k^{(s+1)} +A_j x_j + \sum_{k=j+1}^J A_k x_k^{(s)} - b -\frac{1}{\rho}\Lambda \|^2_2, \ \ j=1,\cdots, J, \label{eq:multi-block-updates-1} \\ 
\Lambda^{(s+1)} & =\Lambda^{(s)}-\rho \biggl(\sum_{j=1}^J A_j x_j^{(s+1)})-b\biggr).\label{eq:multi-block-updates-2}
\end{align}
The first order optimality conditions for \eqref{eq:multi-block-updates-1} are 
\begin{equation}\label{eq:multi-block-optimality-conditions}
g_j(x_j)^{(s+1)} -A^\top_j\Lambda^{(s)} +\rho A_j^\top
\bigl(\sum_{k=1}^j A_k x_k^{(s+1)}  + \sum_{k=j+1}^J A_k x_k^{(s)} -b  \bigr)=0, \ \ j=1,\cdots, J,  \end{equation}
where $g_j(x_j) \in \partial (f_j(x_j)+\mathbf{1}_{\mathcal{X}_j})$ is a subgradient of $(f_j+\mathbf{1}_{\mathcal{X}_j})$ (gradient for differentiable $f_j$'s).

Then, using the update rule \eqref{eq:multi-block-updates-2} of the dual variable, \eqref{eq:multi-block-optimality-conditions} takes the form
\begin{equation}\label{eq:multi-block-optimality-conditions-1}
g_j(x_j)^{(s+1)} -A^\top_j\Lambda^{(s+1)} +\rho A_j^\top
\bigl(\sum_{k=j+1}^J A_k (x_k^{(s)}-x_k^{(s+1)})  \bigr)=0, \ \ j=1,\cdots, J,  
\end{equation}
or equivalently
\begin{equation}\label{eq:multi-block-optimality-conditions-2}
g_j(x_j)^{(s+1)} -A^\top_j\Lambda^{(s+1)}= -\rho A_j^\top
\bigl(\sum_{k=j+1}^J A_k (x_k^{(s)}-x_k^{(s+1)})  \bigr), \ \ j=1,\cdots, J,  
\end{equation}

Further, from the definition of the augmented Lagrangian \eqref{eq:general-augmented-L}, it can easily be seen that for $j=1,\cdots,J$
\begin{equation}\label{eq:subgradient-Lagrangian-1}
g_j(x_j^{(s+1)}) - A^\top_j\Lambda^{(s+1)} +\rho A_j^\top
\bigl(\sum_{k=1}^j A_k x_k^{(s+1)}  + \sum_{k=j+1}^J A_k x_k^{(s+1)} -b  \bigr) \in \partial_{x_j} L_\rho(x_1^{(s+1)},\cdots,x_J^{(s+1)};\Lambda^{(s+1)}), \end{equation}
and
\begin{equation}\label{eq:subgradient-Lagrangian-2}
b  -\sum_{j=1}^J A_j x_j^{(s+1)} \in \partial_{\Lambda} L_\rho(x_1^{(s+1},\cdots,x_J^{(s+1)};\Lambda^{(s+1}))
\end{equation}
Finally, define
\begin{equation}\label{eq:residuals-1}
R_j^{(s+1)} := \rho A_j^\top \bigl(\sum_{j=1}^J A_j x_j^{(s+1)} -b\bigr) + \rho A_j^\top \bigl(\sum_{k=j+1}^J A_j (x_j^{(s)}-x_j^{(s+1)})\bigr), \ \ j=1,\cdots,J,
\end{equation}
and
\begin{equation}\label{eq:residuals-2}
\Phi^{(s+1)} := b-(\sum_{j=1}^J A_j x_j^{(s+1)}\bigr).
\end{equation}
whose properties are established in auxiliary Lemma~\ref{lemma-residuals}.

\paragraph{Back to the proof of Lemma~\ref{lemma-limit-point}.} Note that the augmented Lagrangian function in  \eqref{eq:augm-lagrangian} is of the form in \eqref{eq:general-augmented-L}, where $f_1$ corresponds to
$\ell(A,B;\text{data})$, $f_2$ to $\mu_A\|A\|_1$, $f_3$ to $\mu_B \|B\|_1$, $f_4$ to $\|A-\widetilde{A}\|_\F^2$, $f_5$ to $\|B-\widetilde{B}\|_\F^2$ and $f_6$ to $\sum\nolimits_k\sum\nolimits_{i \neq j}\big( |\widetilde{A}_{ij}|w_{ij} + \tau(1-w_{ij}) + \xi_{ijk} - \tau \lambda_{ik} - \tau \mathbb{I}(j\neq k) + \tau\lambda_{jk}\big)^2$, whereas the constraints are given by $A-\widetilde{A}=0$, $B-\widetilde{B}=0$ and $\sum\nolimits_k\sum\nolimits_{i \neq j}\big( |\widetilde{A}_{ij}|w_{ij} + \tau(1-w_{ij}) + \xi_{ijk} - \tau \lambda_{ik} - \tau \mathbb{I}(j\neq k) + \tau\lambda_{jk}\big)=0$.

With a slight abuse of notation, let $\Theta = (A,\widetilde{A},B,\widetilde{B},\boldsymbol{\lambda},\boldsymbol{\xi}) =: (\Theta_1,\cdots, \Theta_6)$. Then, the terms defined in \eqref{eq:residuals-1} and \eqref{eq:residuals-2} correspond to:
\begin{align*}
R_{\Theta,r}^{(s+1)} & :=\rho\sum_{l=1}^6 \Theta_r^{(s+1)} + \rho\bigl(\sum_{l>r}^6 (\Theta^{(s)}_l-\Theta^{(s+1)}_l)\bigr),~~r = 1,\cdots, 6,\\ 
\Phi_A^{(s+1)} & := -(A^{(s+1)}-\widetilde{A}^{(s+1)}), \\
\Phi_B^{(s+1)} & := -(B^{(s+1)}-\widetilde{B}^{(s+1)}), \\
\Phi_y^{(s+1)} & := -\bigl(\sum\nolimits_k\sum\nolimits_{i \neq j}\big( |\widetilde{A}_{ij}|w_{ij} + \tau(1-w_{ij}) + \xi_{ijk} - \tau \lambda_{ik} - \tau \mathbb{I}(j\neq k) + \tau\lambda_{jk}\bigr).
\end{align*}
Then, the first result of Lemma \ref{lemma-residuals} establishes
that 
\begin{equation}\label{eq:residuals-subdifferential}
\bigl(R_{\Theta,1}^{(s+1)},R^{(s+1)}_{\Theta,2},R_{\Theta,3}^{(s+1)},R^{(s+1)}_{\Theta,4},R_{\Theta,5}^{(s+1)},R_{\Theta,6}^{(s+1)},\Phi_{U_A}^{(s+1)},\Phi_{U_B}^{(s+1)},\Phi_y^{(s+1)}\bigr) \in \partial L_\rho(\Theta^{(s+1)};\Psi^{(s+1)}).
\end{equation}
The second result of Lemma \ref{lemma-residuals} establishes that
\begin{equation*}
\begin{split}
 \sum_{r=1}
R_{\Theta,r}^{(s+1)} + & \sum_{\chi\in\{U_A,U_B,y\}} \Phi_{\chi}^{(s+1)} \leq \\
& \bar{U} \biggl( \| A^{(s)}-A^{(s+1)} \|_\F^2 + \|\widetilde{A}^{(s)}-
\widetilde{A}^{(s+1)}\|_{\F}^2 + \|B^{(s)}-B^{(s+1)}\|_{\F}^2 + \|\widetilde{B}^{(s)} -\widetilde{B}^{(s+1)}\|_{\F}^2 \\
& \quad +\|\boldsymbol{\lambda}^{(s)}-\boldsymbol{\lambda}^{(s+1)}\|_{\F}^2+\|\boldsymbol\xi^{(s)}-\boldsymbol\xi^{(s+1)}\|_{\F}^2 \\
& \quad + \|U_A^{(s)}-U_A^{(s+1)}\|_{\F}^2+\|U_B^{(s)}-U_B^{(s+1)}\|_{\F}^2+\|\boldsymbol y^{(s)}-\boldsymbol y^{(s+1)}\|_F^2 \biggr),
\end{split}
\end{equation*}
where $\bar{U}$ is identically defined to that in Lemma~\ref{lemma-residuals} but with the arguments substituted accordingly. Next, using \eqref{eq:lower-bound-Lagrangian} establishes that
\begin{align}
\lim_{s\rightarrow\infty} &\biggl( \| A^{(s)}-A^{(s+1)} \|_\F^2 + \|\widetilde{A}^{(s)}-
\widetilde{A}^{(s+1)}\|_\F^2 + \|B^{(s)}-B^{(s+1)}\|_\F^2
+ \|\widetilde{B}^{(s)} -\widetilde{B}^{(s+1)}\|_\F^2 + \|\boldsymbol{\lambda}^{(s)}-\boldsymbol{\lambda}^{(s+1)}\|_\F^2 \nonumber \\ 
&+\|\boldsymbol\xi^{(s)}-\boldsymbol\xi^{(s+1)}\|_\F^2 +
\|U_A^{(s)}-U_A^{(s+1)}\|_\F^2+\|U_B^{(s)}-U_B^{(s+1)}\|_\F^2+\|\boldsymbol y^{(s)}- \boldsymbol y^{(s+1)}\|_\F^2 \biggr) = (0,\cdots,0). \label{eq:diff-parameters-0}
\end{align}
In the case where Algorithm~2 does not stop at a stationary point, using Lemma \ref{lemma-subsequence}, there
exists a subsequence $(\Theta^{(s_v)},\Psi^{(s_v)})$, such that 
$\lim_{v\rightarrow \infty} (\Theta^{(s_v)},\Psi^{(s_v)})=
(\Theta^*,\Psi^*)$. Then, using \eqref{eq:residuals-subdifferential} and \eqref{eq:diff-parameters-0}, we obtain that $(0,\cdots,0)\in \partial L_\rho(\Theta^*,\Psi^*)$.
\end{proof}

\begin{lemma}\label{lemma-residuals}
Consider the terms $R_j^{(s+1)}$ and $\Phi^{(s+1)}$ defined in \eqref{eq:residuals-1} and \eqref{eq:residuals-2}, respectively, produced at the $(s+1)$-iteration of a multi-block ADMM algorithm optimizing \eqref{eq:general-augmented-L}, obtained from problem formulation \eqref{eq:general-problem}.
Then, the following hold
\begin{enumerate}
\item $\bigl(R_1^{(s+1)},\cdots,R_J^{(s+1)},\Phi^{(s+1)}\bigr)\in\partial L_\rho(x_1^{(s+1)},\cdots,x_J^{(s+1)};\Lambda^{(s+1)})$;
\item With $\bar{U}\equiv\max\big\{\rho\sum_{j=1}^J \|A_j^\top\|_\F^2,~~\frac{1}{\rho}+1+\sum_{j=1}^J \|A_j^\top\|_\F^2\big\}$,
\begin{equation*}
\|\bigl(R_1^{(s+1)},\cdots,R_J^{(s+1)},\Phi^{(s+1)}\bigr)\|  \leq \bar{U} \|
\sum_{j=1}^J A_j (x_j^{(s)}- x_j^{(s+1)})\|+\|\Lambda^{(s)}-\Lambda^{(s+1)}\|.
\end{equation*}
\end{enumerate}
\end{lemma}

\begin{proof}
For claim 1, substituting \eqref{eq:multi-block-optimality-conditions-2} into \eqref{eq:subgradient-Lagrangian-1} and also the definition of $\Phi^{(s+1)}$ in \eqref{eq:residuals-2} and \eqref{eq:subgradient-Lagrangian-2} yields the result.

For claim 2, we have $\|\bigl(R_1^{(s+1)},\cdots,R_J^{(s+1)},\Phi^{(s+1)}\bigr)\|
\leq \sum_{j=1}^J \|R_j^{(s+1)}\| + \|\Phi^{(s+1)}\|$.
We then bound each term in the latter expression.
\begin{align*}
\|R_j^{(s+1)}\| & \leq \rho \|A_j^\top\|_\F^2 \|\sum_{j=1}^J x_j^{(s+1)}-b\| +
\rho \|A_j^\top\|_\F^2 \bigl( \sum_{j=1}^J \|A_j x_j^{(s)} - 
A_j x_j^{(s+1)} \| \bigr), \\
& \leq \|A_j^\top\|_\F^2 \|\Lambda^{(s)}-\Lambda^{(s+1)}\| +
\rho \|A_j^\top\|_\F^2 \bigl( \sum_{j=1}^J \|A_j x_j^{(s)} - 
A_j x_j^{(s+1)} \| \bigr), \ \ j=1,\cdots,J
\end{align*}
and
\begin{equation*}
\|\Phi^{(s+1)}\| =  \|\sum_{j=1}^J x_j^{(s+1)}-b\|=\frac{1}{\rho} 
 \|\Lambda^{(s)}-\Lambda^{(s+1)}\|.
\end{equation*}
Therefore, 
\begin{equation*}
    \sum_{j=1}^J \|R_j^{(s+1)}\| + \|\Phi^{(s+1)}\| \leq \bar{U}
    \|\sum_{j=1}^J A_j (x_j^{(s)}-x_j^{(s+1)}) \|+ 
    \|\Lambda^{(s)}-\Lambda^{(s+1)}\|.
\end{equation*}

\end{proof}

%% file: A3_background_id.tex
\section{Identification for a low-dimensional SVAR}\label{sec:background-id}

We provide a brief discussion of a widely used identification scheme in the econometrics literature for low dimensional SVAR models.

The SVAR model in~(1) can be written in a reduced VAR form as
$$X_t=\sum_{j-1}^d (\mathrm{I}_p-A)^{-1} B_j + (\mathrm{I}_p-A)^{-1}\epsilon_t.$$ Then, the covariance matrix of $u_t :=(\mathrm{I}_p-A)^{-1}\epsilon_t$ is given by
$\Sigma_u=(\mathrm{I}_p-A)^{-1}\Sigma_\epsilon (\mathrm{I}_p-A)^{-1}$. Since the model is low dimensional, the elements of $\Sigma_u$ can be estimated consistently from the data. Note that the number of free parameters is $(p^2+p)$ ($p^2$ in $A$ and $p$ in the diagonal $\Sigma_\epsilon$), while the number of unique equations in $(\mathrm{I}_p-A)^{-1}\Sigma_\epsilon (\mathrm{I}_p-A)^{-1}$ are $p(p+1)/2$ due to the symmetry of the covariance matrix. Hence, identification of $A$ and $\Sigma_\epsilon$ requires $p(p+1)/2$ additional constraints. Customarily, $p$ of those come by assuming that the diagonal elements of $\Sigma_\epsilon$ are equal to 1. The remaining $p(p-1)/2$ constraints come either by specifying certain entries of $A$, or by assuming that entries in $A$ are proportional to other entries; both such assumptions in econometric applications come from theoretical economic models, or other prior domain knowledge \citep{stock2016dynamic}. Since in many applications $p$ ranges from 2-5, it is feasible to come up with the required number of constraints.

%% file: A4_numerical.tex
\section{Additional Notes for Numerical Experiments}\label{appendix:sim}

We provide additional notes for numerical experiments. In Sections~\ref{appendix:competitor} and~\ref{sec:varsortability}, based on synthetic data settings S1-S4, we present performance evaluation for several additional competing methods in the absence of the partial ordering, with the discussions focusing on the structural components; in Section~\ref{appendix:lag}, we present performance evaluation for the recovered lag components as well as the overall goodness-of-fit for the underlying time series for the experiments considered in Section~\ref{sec:simulation}.

\subsection{Competing methods in the absence of partial ordering}\label{appendix:competitor}
We consider the following competing models, described below:
\begin{itemize}[topsep=0pt,itemsep=0pt]
    \item Dynotears \citep{pamfil2020dynotears}, where the structural parameter and lag parameters are modelled jointly, leveraging the Notears algorithm \citep{zheng2018dags}
    \item NotearsLasso, where a two-stage procedure is implemented: at stage 1, use Notears \citep{zheng2018dags} to recover the structural component $A$ according to an SEM representation $X_t = AX_t + v_t$, in the absence of lag terms; at stage 2, after obtaining the residuals, that is, $\widehat{v}_t := x_t - \widehat{A}x_t$, perform Lasso regression \citep{tibshirani1996regression} with $v_t$ as the response and the lags $x_{t-\ell}$ as the predictors.
    \item tsLiNGAM \citep{hyvarinen2010estimation}, where one starts from the reduced form representation, that is $X_t = \sum \tilde{B}_{\ell}X_{t-\ell}+u_t$ and estimate the parameters $\tilde{B}_\ell$'s with Lasso, then perform LiNGAM \citep{shimizu2006linear} on the residuals $\widehat{u}_t := x_t - \sum \widehat{\tilde{B}}_{\ell}x_{t-\ell}$, to obtain $\widehat{A}$; finally, $\widehat{B}_{\ell}:= (\mathrm{I}_p - \widehat{A})\widehat{\tilde{B}}_{\ell}$
\end{itemize}
Table~\ref{tab:simres-competitor} displays the evaluation results for the structural parameter across synthetic data settings S1-S4. We note that Dynotears has the strongest performance across all settings; the NotearsLasso exhibits very decent discovery, although the true negative rate does not improve much despite increasing sample size. For tsLiNGAM, it suffers from low detection despite the presence of an adequate sample size.  
\begin{table*}[h]
\scriptsize
\setstretch{0.8}
\centering
\caption{Evaluation for $\widehat{A}$ obtained from competing models, in the \textit{absence} of prior information}\label{tab:simres-competitor}\vspace*{-2mm}
\begin{tabular}{rr|cc|cc|cc}
\toprule
&  & \multicolumn{2}{c|}{Dynotears} & \multicolumn{2}{c|}{NotearsLasso} & \multicolumn{2}{c}{tsLiNGAM} \\ \cmidrule(lr){3-4} \cmidrule(lr){5-6} \cmidrule(lr){7-8}
& $n$ & TP & TN & TP & TN & TP & TN \\ \hline
S1 & 50 & 0.74(.03) & 0.80(.004) & 0.73(.03) & 0.73(.005) & 0.19(.03) & 0.99(.001)\\
& 100 & 0.82(.03) & 0.95(.002) &  0.87(.03) & 0.82(.005) & 0.32(.04) & 0.98(.002) \\
& 200 & 0.90(.01) & 0.97(.002) & 0.93(.007) & 0.83(.006) & 0.45(.04) & 0.97(.002) \\ \midrule
S2 & 50 & 0.65(.03) & 0.80(.004) & 0.62(.03) & 0.71(.007) & 0.15(.02) & 0.98(.002)\\
& 100 & 0.78(.03) & 0.94(.003) & 0.79(.03) & 0.72(.006) & 0.20(.02) & 0.97(.001) \\
& 200 & 0.92 (.01) & 0.95(.002)& 0.83(.02) & 0.76(.004) & 0.32(.02) & 0.97(.002) \\ \midrule
S3 & 50 & 0.73(.04) & 0.81(.004) & 0.72(.03) & 0.74(.005) &  0.22(.02) & 0.99(.001) \\
& 100 & 0.83(.03) & 0.96(.002) & 0.86(.02) & 0.84(.005) & 0.37(.04) & 0.98(.001) \\ 
& 200 & 0.92(.01) & 0.98(.001) & 0.90(.02) & 0.85(.004) & 0.55(.05) & 0.98(.001) \\ \midrule
S4 & 50 & 0.61(.03) & 0.85(.003) & 0.60(.02) & 0.71(.01) & 0.11(.02) & 0.98(.003) \\
& 100 & 0.83(.01) & 0.94(.001) & 0.78(.02) & 0.72(.008) & 0.22(.02) & 0.98(.002) \\
& 200 & 0.90(.02) & 0.96(.001) & 0.82(.02) & 0.77(.007) & 0.38(.02) & 0.96(.002) \\ 
\bottomrule
    \end{tabular}
\end{table*}

\subsection{Varsortability, data normalization and performance}\label{sec:varsortability}

All experimental results presented thus far are based on unnormalized data. In \citet{reisach2021beware}, the authors report that the performance of DAG estimation using continuous structural methods (e.g., NOTEARS, \citep{zheng2018dags}) can be susceptible to data normalization; see also discussion in \citet{kaiser2022unsuitability}. In particular, \citet{reisach2021beware} define ``varsortability" $v$ as a measure of the causal structure information embedded in the data scale; when $v=1$, the causal structure can be recovered by ordering the nodes with increasing marginal variance. In their benchmarking scenario, varsortability of simulated linear SEMs with unnormalized data averages around 0.94, whereas after data normalization around 0.5. 

In light of the above papers, we examine the varsortability for our simulated data, which averages at 0.96. Given that the benchmarking methods (e.g., Dynotears and NotearsLasso) are time-series extensions of NOTEARS, it's likely that they are benefiting from the high varsortability embedded in the data. See, e.g., Table~\ref{tab:dynotears-normal} where the same set of experiments are conducted using Dynotears with $n=200$ on normalized data; a noticeable impact from normalization is that the true positive rate deteriorates and it drops dramatically for settings S2 and S4 where the graph is denser:\footnote{Tuning parameter is chosen such that TN stays in the same ballpark as the case where data is unnormalized. In the case where one chooses a different set of tuning parameters (in particular, for S2 and S4) such that TN is more comparable to the ones shown in Table~\ref{tab:simres-normalize}, TP would be in the range of high 30\% and low 40\%.}
\begin{table}[!h]
\centering
\scriptsize
\setstretch{0.8}
\caption{Evaluation for $\widehat{A}$ for Dynotears, after data normalization; $n=200$. Results are based on the median of 10 replicates, with the standard deviation reported in parentheses.}\label{tab:dynotears-normal}\vspace*{-2mm}
\begin{tabular}{c|cc|cc|cc|cc}
\toprule
    & \multicolumn{2}{c|}{S1} & \multicolumn{2}{c|}{S2} & \multicolumn{2}{c|}{S3} & \multicolumn{2}{c}{S4} \\  \cmidrule(lr){2-3} \cmidrule(lr){4-5} \cmidrule(lr){6-7} \cmidrule(lr){8-9}
    & TP & TN & TP & TN & TP & TN & TP & TN\\ \hline
 $n=200$  &  0.84 (.03) & 0.97 (.001) & 0.27 (.05) & 0.96 (.003) & 0.82 (.03) & 0.96 (.001) & 0.25 (.05) & 0.96 (.001) \\
 \bottomrule
\end{tabular}
\end{table}

Since the method adopted in this work is also an optimization-based one, we would like to further investigate whether the same issue stands. To this end, additional experiments are conducted for our proposed method based on normalized data, focusing on the case where sample size $n=200$. 
\begin{table}[!ht]
\scriptsize
\setstretch{0.8}
\centering
\caption{Evaluation for $\widehat{A}$ of our proposed method after data normalization; $n=200$. Results are based on the median of 10 replicates, with the standard deviation reported in parentheses.}\label{tab:simres-normalize}\vspace*{-2mm}
\begin{tabular}{r|cc|cc|cc|cc}
\toprule
&  \multicolumn{2}{c|}{00} &  \multicolumn{2}{c|}{10} & \multicolumn{2}{c|}{20} & \multicolumn{2}{c}{50} \\ \cmidrule(lr){2-3} \cmidrule(lr){4-5} \cmidrule(lr){6-7} \cmidrule(lr){8-9}
& TP & TN & TP & TN & TP & TN & TP & TN\\ \hline
S1 & 0.93(0.009) & 0.96(0.002) & 0.93(0.011) & 0.96(0.002) & 0.94(0.011) & 0.97(0.002) & 0.96(0.011) & 0.98(0.002) \\

S2 & 0.84(0.047) & 0.84(0.008) & 0.84(0.046) & 0.85(0.008) & 0.85(0.044) & 0.86(0.006) & 0.91(0.048) & 0.88(0.005) \\

S3 & 0.96(0.011) & 0.96(0.004) & 0.96(0.012) & 0.96(0.004) & 0.96(0.012) & 0.96(0.004) & 0.98(0.007) & 0.97(0.003) \\

S4 & 0.86(0.055) & 0.82(0.014) & 0.86(0.054) & 0.83(0.013) & 0.86(0.053) & 0.84(0.013) & 0.91(0.045) & 0.85(0.008) \\

S5 & 0.96(0.010) & 0.96(0.004) & 0.96(0.010) & 0.96(0.004) & 0.96(0.010) & 0.96(0.003) & 0.98(0.007) & 0.97(0.003) \\

S6 & 0.82(0.060) & 0.85(0.013) & 0.82(0.054) & 0.86(0.012) & 0.82(0.055) & 0.87(0.011) & 0.88(0.049) & 0.89(0.006) \\
\hline
\end{tabular}
\end{table}

Observations based on the above experimental results are two-fold: (i) for settings S1, S3 and S5 that are more sparse---irrespective of the noise distribution---standardization actually improves the performance of the proposed algorithm; e.g., in the absence of partial ordering information, both TP and TN go up to 90+\% (without standardization the same metric is in the high 80\%); and (2) for settings S2, S4 and S6 where the graphs are denser, the performance is largely unaffected.

One explanation of these experimental findings is as follows. Based on the argument in \cite{reisach2021beware}, varsortability positively affects continuous structure learning methods through the gradient of the MSE-based score functions (Section 3.4). In particular, the optimization procedure leveraged in these methods relies on the gradient information; through the interplay between gradient and step size that is optimized by line-search, high varsortability results in preference for causal edges (Section E.5). The method developed in the current paper also falls under the realm of continuous structure learning methods. However, when solving the objective function via ADMM, each primal block update possesses a closed-form solution; specifically, the structural component (i.e., $A$) is obtained by solving a linear system and thus there is no gradient step that drives the estimates of $A$ (within any single iteration). In other words, although the proposed algorithm optimizes an MSE-based score function, the inner workings of ADMM operate in a way such that the benefit of high varsortability is essentially muted, since there are no explicit gradient steps that directly facilitate the better calibration of high-marginal variance nodes.

\subsection{Performance evaluation for lag components and the overall goodness-of-fit}\label{appendix:lag}

We present the performance of the skeleton recovery of the lag components for settings S1-S6 in Tables~\ref{tab:B1} and~\ref{tab:B2}. Recall that across all settings, two lags are included in the DGP and the density levels (i.e., the cardinality of the support set over $p^2$, the total number of edges in the fully-connected case) of $B_1$ and $B_2$ are fixed at 5\% and 2\%, respectively. 

\begin{table*}[!ht]
\scriptsize
\setstretch{0.9}
\centering
\caption{Evaluation for $\widehat{B}_1$ obtained using our proposed method. Results are based on the median of 10 replicates, with the standard deviation of the corresponding metric reported in parentheses.}\label{tab:B1}\vspace*{-2mm}
\begin{tabular}{rr|cc|cc|cc|cc}
\toprule
&  & \multicolumn{2}{c|}{00} &  \multicolumn{2}{c|}{10} & \multicolumn{2}{c|}{20} & \multicolumn{2}{c}{50} \\ \cmidrule(lr){3-4} \cmidrule(lr){5-6} \cmidrule(lr){7-8} \cmidrule(lr){9-10}
& $n$ & TP & TN & TP & TN & TP & TN & TP & TN\\ \hline

S1 & 100  & 0.60(0.013) & 0.79(0.007) & 0.62(0.015) & 0.79(0.007) & 0.62(0.013) & 0.79(0.008) & 0.65(0.017) & 0.78(0.006) \\
   & 200 & 0.79(0.007) & 0.82(0.006) & 0.80(0.010) & 0.82(0.007) & 0.81(0.012) & 0.82(0.007) & 0.82(0.014) & 0.82(0.009) \\\midrule

S2 & 100  & 0.60(0.017) & 0.69(0.004) & 0.60(0.021) & 0.69(0.003) & 0.63(0.032) & 0.67(0.003) & 0.65(0.015) & 0.67(0.007) \\
   & 200 & 0.82(0.020) & 0.84(0.004) & 0.83(0.017) & 0.84(0.005) & 0.85(0.012) & 0.83(0.006) & 0.87(0.013) & 0.84(0.005) \\\midrule

S3 & 100  & 0.63(0.012) & 0.68(0.005) & 0.65(0.013) & 0.67(0.005) & 0.66(0.009) & 0.67(0.006) & 0.67(0.017) & 0.66(0.005) \\
   & 200 & 0.84(0.016) & 0.82(0.002) & 0.85(0.015) & 0.82(0.003) & 0.86(0.014) & 0.82(0.003) & 0.88(0.012) & 0.82(0.003) \\\midrule

S4 & 100  & 0.62(0.024) & 0.66(0.005) & 0.63(0.026) & 0.65(0.006) & 0.65(0.025) & 0.64(0.004) & 0.68(0.020) & 0.63(0.005) \\
   & 200 & 0.82(0.007) & 0.86(0.004) & 0.84(0.013) & 0.86(0.003) & 0.85(0.010) & 0.86(0.004) & 0.88(0.010) & 0.87(0.004) \\\midrule

S5 & 100  & 0.61(0.021) & 0.78(0.002) & 0.63(0.023) & 0.77(0.004) & 0.64(0.027) & 0.77(0.004) & 0.66(0.025) & 0.76(0.006) \\
   & 200 & 0.82(0.012) & 0.83(0.006) & 0.84(0.013) & 0.83(0.004) & 0.84(0.012) & 0.83(0.005) & 0.86(0.015) & 0.83(0.004) \\\midrule

S6 & 100  & 0.62(0.017) & 0.76(0.004) & 0.64(0.011) & 0.75(0.004) & 0.66(0.014) & 0.75(0.005) & 0.69(0.011) & 0.74(0.004) \\
   & 200 & 0.85(0.008) & 0.84(0.005) & 0.86(0.009) & 0.84(0.005) & 0.86(0.010) & 0.84(0.004) & 0.89(0.005) & 0.85(0.003) \\
\bottomrule
\end{tabular}
\caption{Evaluation for $\widehat{B}_2$ obtained using our proposed method. Results are based on the median of 10 replicates, with the standard deviation of the corresponding metric reported in parentheses.}\label{tab:B2}
\begin{tabular}{rr|cc|cc|cc|cc}
\toprule
&  & \multicolumn{2}{c|}{00} &  \multicolumn{2}{c|}{10} & \multicolumn{2}{c|}{20} & \multicolumn{2}{c}{50} \\ \cmidrule(lr){3-4} \cmidrule(lr){5-6} \cmidrule(lr){7-8} \cmidrule(lr){9-10}
& $n$ & TP & TN & TP & TN & TP & TN & TP & TN\\ \hline

S1 & 100  & 0.65(0.021) & 0.78(0.007) & 0.66(0.025) & 0.78(0.007) & 0.67(0.026) & 0.77(0.006) & 0.71(0.026) & 0.77(0.005) \\
   & 200 & 0.85(0.020) & 0.81(0.010) & 0.85(0.019) & 0.81(0.009) & 0.85(0.018) & 0.81(0.009) & 0.86(0.013) & 0.81(0.009) \\\midrule

S2 & 100  & 0.66(0.038) & 0.66(0.006) & 0.66(0.032) & 0.65(0.005) & 0.67(0.036) & 0.64(0.006) & 0.70(0.028) & 0.63(0.004) \\
   & 200 & 0.88(0.020) & 0.82(0.007) & 0.88(0.020) & 0.82(0.006) & 0.88(0.020) & 0.82(0.006) & 0.90(0.018) & 0.82(0.006) \\\midrule

S3 & 100  & 0.69(0.028) & 0.66(0.004) & 0.70(0.032) & 0.65(0.004) & 0.69(0.029) & 0.65(0.005) & 0.71(0.032) & 0.64(0.004) \\
   & 200 & 0.87(0.010) & 0.82(0.004) & 0.87(0.011) & 0.82(0.005) & 0.87(0.013) & 0.82(0.005) & 0.88(0.014) & 0.82(0.004) \\\midrule

S4 & 100  & 0.67(0.028) & 0.63(0.005) & 0.68(0.030) & 0.61(0.004) & 0.69(0.043) & 0.60(0.005) & 0.72(0.033) & 0.60(0.005) \\
   & 200 & 0.88(0.021) & 0.85(0.003) & 0.89(0.027) & 0.85(0.003) & 0.90(0.022) & 0.85(0.004) & 0.92(0.016) & 0.85(0.003) \\\midrule

S5 & 100  & 0.68(0.030) & 0.76(0.004) & 0.68(0.023) & 0.75(0.004) & 0.70(0.022) & 0.75(0.002) & 0.71(0.023) & 0.74(0.005) \\
   & 200 & 0.86(0.019) & 0.82(0.004) & 0.86(0.020) & 0.82(0.004) & 0.87(0.028) & 0.82(0.004) & 0.89(0.027) & 0.82(0.003) \\\midrule

S6 & 100  & 0.69(0.039) & 0.73(0.003) & 0.70(0.046) & 0.72(0.004) & 0.72(0.035) & 0.72(0.005) & 0.73(0.036) & 0.72(0.004) \\
   & 200 & 0.90(0.022) & 0.83(0.004) & 0.90(0.019) & 0.83(0.003) & 0.91(0.019) & 0.83(0.003) & 0.93(0.010) & 0.83(0.004) \\
\bottomrule
\end{tabular}
\end{table*}

Observations based on the results in Tables~\ref{tab:B1} and~\ref{tab:B2} are two-fold: (1) as sample size increases, the skeleton recovery performance of $B$'s improves with a more pronounced increase in the true-positive-rate (TP); and (2) the recover of $B_2$ is superior than that of $B_1$, and this is consistent with the fact that the underlying true $B_2$ has a lower density level than $B_1$, and therefore the recovery of the former is generally an easier task than that of the latter (see also Remark~\ref{rmk:density}). Finally, we note that in a reduced VAR setting, i.e., when the model does not have a structural component and is given by $X_t = B_1X_{t-1} + B_2 X_{t-2} + \epsilon_t$, with the same sample size, the recovery of the $B$s can typically be better. However, the inferior performance in the current ``structural" setting is somewhat expected, as any estimation error in the structural component will propagate and affect the estimation of the $B$s; in other words, from the standpoint of $B$ estimation, in the structural setting, one is effectively using contaminated data and therefore larger error is expected vis-a-vis the ``clean" case.   

To measure the overall goodness-of-fit, we consider using the relative $\ell_2$ error of the one-step-ahead forecast. Recall that the SVAR model can be written in the form of a reduced VAR (see Section~\ref{sec:formulation}); given observed data $\{x_1,\cdots,x_n\}$, once the model is trained, one can obtain a {\em forecast} $\widehat{x}_{n+1}$ by plugging in the estimated parameters into~\eqref{eq:reduced-var}. The relative $\ell_2$ error is then calculated as $\|\widehat{x}_{n+1}-x_{n+1}\|/\|x_{n+1}\|$. Results are reported in Table~\ref{tab:x}. 

\begin{table*}[h]
\scriptsize
\setstretch{0.9}
\centering
\caption{Relative $\ell_2$ error for the one-step-ahead forecast $\widehat{x}_{n+1}$. Results are based on the mean of 10 replicates, with the standard deviation of the corresponding metric reported in parentheses.}\label{tab:x}
\begin{tabular}{rr|c|c|c|c}
\toprule
&  & 00 & 10 & 20 & 50 \\ \hline
S1 & 100  & 0.07(0.021) & 0.07(0.018) & 0.07(0.016) & 0.07(0.016) \\
   & 200 & 0.06(0.013) & 0.06(0.012) & 0.05(0.012) & 0.05(0.012) \\\midrule

S2 & 100  & 0.05(0.022) & 0.04(0.014) & 0.04(0.009) & 0.03(0.009) \\
   & 200 & 0.03(0.007) & 0.03(0.006) & 0.03(0.006) & 0.02(0.007) \\\midrule

S3 & 100  & 0.09(0.017) & 0.09(0.021) & 0.08(0.017) & 0.08(0.020) \\
   & 200 & 0.06(0.014) & 0.06(0.014) & 0.06(0.013) & 0.05(0.013) \\\midrule

S4 & 100  & 0.07(0.042) & 0.05(0.015) & 0.05(0.014) & 0.04(0.010) \\
   & 200 & 0.02(0.009) & 0.02(0.010) & 0.02(0.010) & 0.02(0.009) \\\midrule

S5 & 100  & 0.07(0.011) & 0.07(0.013) & 0.07(0.011) & 0.07(0.012) \\
   & 200 & 0.05(0.010) & 0.06(0.011) & 0.06(0.010) & 0.05(0.009) \\\midrule

S6 & 100  & 0.06(0.025) & 0.04(0.015) & 0.04(0.015) & 0.04(0.013) \\
   & 200 & 0.02(0.008) & 0.02(0.006) & 0.02(0.006) & 0.02(0.006) \\
\bottomrule
\end{tabular}
\end{table*}

Based on the reported metrics, the overall goodness-of-fit is decent, with the relative $\ell_2$ error below 10\% irrespective of the prior settings. Note that although the higher-density settings (e.g., S2 vs. S1) typically yield worse skeleton recovery performance for both the structural and the lag components, from a pure forecasting perspective, such settings produce smaller errors. Additionally, the impact of prior information on the skeleton is more pronounced in the case where the density level of the structural component is high and the sample size is relative small.

%% file: A5_listofVariables.tex
\section{Variables Info for Analyzing US Macroeconomic Indicators}

{\scriptsize
\setstretch{0.8}
\begin{longtable}{llr}
\caption{List of US macroeconomic indicators used in Section~5, their descriptions and tiers.}\\
\hline
MNEMONIC &DESCRIPTION & TIER\\ \hline
DPIC96 &Real Disposable Personal Income (Billions of Chained 2012 Dollars) & 1\\
INDPRO &Industrial Production Index (Index 2012=100) & 1\\
IPFINAL &IP: Final Products (Market Group) (Index 2012=100) & 1\\
IPCONGD &IP: Consumer Goods (Index 2012=100) & 1\\
IPMAT &IP: Materials (Index 2012=100) & 1\\
IPDMAT &IP: Durable Materials (Index 2012=100) & 1\\
IPNMAT &IP: Nondurable Materials (Index 2012=100) & 1\\
IPDCONGD &IP: Durable Consumer Goods (Index 2012=100) & 1\\
IPB51110SQ &IP: Durable Goods: Automotive products (Index 2012=100) & 1\\
IPNCONGD &IP: Nondurable Consumer Goods (Index 2012=100) & 1\\
IPBUSEQ &IP: Business Equipment (Index 2012=100) & 1\\
IPB51220SQ &IP: Consumer energy products (Index 2012=100) & 1\\
TCU &Capacity Utilization: Total Industry (Percent of Capacity) & 1\\
PAYEMS &All Employees: Total nonfarm (Thousands of Persons) & 1\\
USPRIV &All Employees: Total Private Industries (Thousands of Persons) & 1\\
USGOVT &All Employees: Government (Thousands of Persons) & 1\\
CIVPART &Civilian Labor Force Participation Rate (Percent) & 1\\
UNRATE &Civilian Unemployment Rate (Percent) & 1\\
PERMIT &New Private Housing Units Authorized by Building Permits (Thousands of Units) & 3\\
HOUSTMW &Housing Starts in Midwest Census Region (Thousands of Units) & 3\\
HOUSTNE &Housing Starts in Northeast Census Region (Thousands of Units) & 3\\
HOUSTS &Housing Starts in South Census Region (Thousands of Units) & 3\\
HOUSTW &Housing Starts in West Census Region (Thousands of Units) & 3\\
AMDMUOx & \pbox[t]{15cm}{Real Value of Manufacturers' Unfilled Orders for Durable Goods Industries, \\ deflated by Core PCE}  & 3\\
ANDENOx & \pbox[t]{15cm}{Real Value of Manufacturers' New Orders for Capital Goods:\\
Nondefense Capital Goods Industries, deflated by Core PCE}  & 3\\
INVCQRMTSPL &Real Manufacturing and Trade Inventories (Millions of 2012 Dollars) & 3\\
PCECTPI &Personal Consumption Expenditures: Chain-type Price Index (Index 2009=100) & 1\\
DGDSRG3Q086SBEA &Personal consumption expenditures: Goods (chain-type price index) & 1\\
DSERRG3Q086SBEA &Personal consumption expenditures: Services (chain-type price index) & 1\\
DNDGRG3Q086SBEA &Personal consumption expenditures: Nondurable goods (chain-type price index) & 1\\
WPSFD49502 &PPI by Commodity for Finished Consumer Goods (Index 1982=100) & 1\\
PPIIDC &PPI by Commodity Industrial Commodities (Index 1982=100) & 1\\
WPSID61 & PPI by Commodity Intermediate Materials: Supplies \& Components (Index 1982=100) & 1\\
WPU0531 & \pbox[t]{15cm}{PPI by Commodity for Fuels and Related Products and Power:\\
Natural Gas (Index 1982=100)} & 1\\
WPU0561 & \pbox[t]{15cm}{PPI by Commodity for Fuels and Related Products and Power:\\ 
Crude Petroleum (Domestic Production) (Index 1982=100)} & 1\\
CES2000000008x & \pbox[t]{15cm}{Real Average Hourly Earnings of Production and Nonsupervisory Employees:\\ Construction, deflated by Core PCE} & 1\\
CES3000000008x & \pbox[t]{15cm}{Real Average Hourly Earnings of Production and Nonsupervisory Employees:\\ Manufacturing, deflated by Core PCE} & 1\\
FEDFUNDS &Effective Federal Funds Rate (Percent) & 2\\
TB3MS &3-Month Treasury Bill: Secondary Market Rate (Percent) & 3\\
TB6MS &6-Month Treasury Bill: Secondary Market Rate (Percent) & 3\\
GS1 &1-Year Treasury Constant Maturity Rate (Percent) & 3\\
GS10 &10-Year Treasury Constant Maturity Rate (Percent) & 3\\
AAA &Moody's Seasoned Aaa Corporate Bond Yield© (Percent) & 3\\
BAA &Moody's Seasoned Baa Corporate Bond Yield© (Percent) & 3\\
TB6M3Mx &6-Month Treasury Bill Minus 3-Month Treasury Bill, secondary market (Percent) & 3\\
GS1TB3Mx &1-Year Treasury Constant Maturity Minus 3-Month Treasury Bill,
secondary market (Percent)& 3\\
GS10TB3Mx &10-Year Treasury Constant Maturity Minus 3-Month Treasury Bill,
secondary market (Percent) & 3\\
BOGMBASEREALx &St. Louis Adjusted Monetary Base (Billions of 1982-84 Dollars), deflated by CPI & 3\\
M1REAL &Real M1 Money Stock (Billions of 1982-84 Dollars), deflated by CPI & 3\\
M2REAL &Real M2 Money Stock (Billions of 1982-84 Dollars), deflated by CPI & 3\\
BUSLOANSx &\pbox[t]{15cm}{Real Commercial and Industrial Loans, All Commercial Banks\\
(Billions of 2009 U.S. Dollars), deflated by Core PCE}  & 3\\
TOTALSLx &Total Consumer Credit Outstanding, deflated by Core PCE  & 3\\
EXSZUSx &Switzerland / U.S. Foreign Exchange Rate & 3\\
EXJPUSx &Japan / U.S. Foreign Exchange Rate & 3\\
EXUSUKx &U.S. / U.K. Foreign Exchange Rate & 3\\
EXCAUSx &Canada / U.S. Foreign Exchange Rate & 3\\
UEMPMEAN &Average (Mean) Duration of Unemployment (Weeks) & 1\\
WPSID62 &PPI: Crude Materials for Further Processing (Index 1982=100) & 1\\
PPICMM & \pbox[t]{15cm}{PPI: Commodities: Metals and metal products:\\
Primary nonferrous metals (Index 1982=100)} & 1\\
CPIAPPSL &CPI for All Urban Consumers: Apparel (Index 1982-84=100) & 1\\
CPITRNSL &CPI for All Urban Consumers: Transportation (Index 1982-84=100) & 1\\
CPIMEDSL &CPI for All Urban Consumers: Medical Care (Index 1982-84=100) & 1\\
CUSR0000SAC &CPI for All Urban Consumers: Commodities (Index 1982-84=100) & 1\\
CUSR0000SAD &CPI for All Urban Consumers: Durables (Index 1982-84=100) & 1\\
CUSR0000SAS &CPI for All Urban Consumers: Services (Index 1982-84=100) & 1\\
CPIULFSL &CPI for All Urban Consumers: All Items Less Food (Index 1982-84=100) & 1\\
CUSR0000SA0L2 &CPI for All Urban Consumers: All items less shelter (Index 1982-84=100) & 1\\
CUSR0000SA0L5 &CPI for All Urban Consumers: All items less medical care (Index 1982-84=100) & 1\\
HWIURATIOx &Ratio of Help Wanted/No. Unemployed & 1\\
CLAIMSx &Initial Claims & 1\\
NASDAQCOM &NASDAQ Composite (Index Feb 5, 1971=100) & 3\\
S\&P 500 &S\&P's Common Stock Price Index: Composite & 3\\
S\&P: indust &S\&P's Common Stock Price Index: Industrials & 3\\
S\&P div yield &S\&P's Composite Common Stock: Dividend Yield & 3\\
S\&P PE ratio &S\&P's Composite Common Stock: Price-Earnings Ratio & 3\\
\hline
\end{longtable}
}